\documentclass[vecphys, multicol]{svmult}
\usepackage{makeidx}         
\usepackage{graphicx}        
\usepackage{multicol}        
\usepackage[bottom]{footmisc}

\makeindex             

\begin{document}
\title*{Massive stars as thermonuclear reactors and their explosions following core collapse 
}
\titlerunning{Thermonuclear fusion and collapse and explosion in massive stars}
\author{Alak Ray}
\authorrunning{A. Ray}
\institute{Tata Institute of Fundamental Research, Mumbai 400 005, India
\texttt{akr@tifr.res.in}
}
\maketitle
\begin{abstract}
Nuclear reactions transform atomic nuclei
inside stars. This is the process of stellar nucleosynthesis.
The basic concepts of determining nuclear reaction rates
inside stars  are reviewed. How stars manage to burn their fuel so slowly most
of the time are also considered. Stellar thermonuclear reactions
involving protons in hydrostatic burning are
discussed first. Then I discuss triple alpha reactions in the helium burning stage.
Carbon and oxygen survive in red giant stars because of the nuclear 
structure of oxygen and neon. Further nuclear burning
of carbon, neon, oxygen and silicon in quiescent conditions
are discussed next. In the subsequent core-collapse phase, 
neutronization due to electron capture from the top of the Fermi sea in
a degenerate core takes place. The expected
signal of neutrinos from a nearby supernova is calculated.
The supernova often explodes
inside a dense circumstellar medium, which is established due to the progenitor star
losing its outermost
envelope in a stellar wind or mass transfer in a binary system. 
The nature of the circumstellar medium and the ejecta of the supernova and their
dynamics are revealed by
observations in the optical, IR, radio, and X-ray bands, and I discuss some
of these observations and their interpretations. 
\end{abstract}
\section{Introduction}
\label{sec:1}
The sun is not commonly considered a star and few would think of 
stars as nuclear reactors. Yet, that {\it is} the way it is, and even our
own world is made out of the ``fall-out"
from stars that blew up and spewed radioactive debris into
the nascent solar system\footnote{Lecture Notes on Kodai School on 
Synthesis of Elements in Stars; eds. Aruna Goswami \& Eswar Reddy, 
Springer Verlag, 2009}.

Nuclear Astrophysics is the field concerning ``the synthesis and
Evolution of atomic nuclei, by thermonuclear reactions, from
the Big Bang to the present. What is the origin of the
matter of which we are made?"\cite{Arn96}.
Our high entropy universe resulting from the Big Bang,
contains many more photons per particle of matter with mass,
e.g. protons and neutrons. Because of the high entropy
as the universe expanded, there was time to manufacture elements
only upto helium and the major products of
cosmic nucleosynthesis remained the light elements hydrogen and helium\footnote
{Note however suggestions \cite{Car84,Car94}
that early generation of stars called Pop III objects
can also contribute to the abundance of $^4He$ seen in the universe today
and the entire helium may not be a product of big
bang nucleosynthesis alone.}.
Stars formed
from this primordial matter. They used these elements, hydrogen and
helium  as fuel
to generate energy like a giant nuclear 
reactor\footnote{Our sun is slowly burning hydrogen into
helium and is presently not exactly the same when it just 
started burning hydrogen. It will appear different once
it exhausts all hydrogen it {\it can} burn in its core. In other
words, nuclear reactions in stellar interiors determine
the life-cycle of stars, apart from providing them with internal
power for heat and light and manufacturing all the heavier elements
that the early universe could not.}. In the process,
the stars could shine and manufacture
higher atomic number elements like carbon, oxygen, calcium and iron
which we and our world are made of. The heavy elements are either dredged up
from the core of the star to the surface of the star from which they
are dispersed by stellar wind or directly ejected into the interstellar
medium when a (massive) star explodes. The stardust is the source of
heavy elements for new generation of stars and sun-like systems.

Our sun is {\it not} a massive star. It burns hydrogen in a set of nuclear reactions
called the pp-chain, whereas the more massive stars {\it presently}
burn their hydrogen by the
so-called CNO-cycle\footnote{Note however that
CN cycle may have driven an early stage convection in the young Sun.}. 
Nevertheless, to put nuclear reactions in stars in perspective,
we shall start with a discussion of how these reactions 
proceed within the sun. There is a correspondence between
the evolutionary state of a star, its external appearance\footnote{Astronomers classify stars
according to their colors and (absorption) line spectra and luminosities. 
Meghnad Saha showed \cite{Sah20,Sah21}
the link between the classification scheme and temperature (and thermal ionization) of 
stellar atmosphere.} and internal
core conditions and the nuclear fuel it burns, -- a sort of a mapping
between the astronomers Hertzsprung-Russel diagram and the
nuclear physicist's chart of the nuclide \cite{Ree68}, until nuclear
burning takes place on too rapid a time scale.

The problem of evolution of stars to their explosion and subsequent
interaction with the circumstellar medium has many time scales (ranging from 
tens of milliseconds to tens of thousands of years) and macroscopic length scales
(from dimensions effectively that of a white dwarf to that of a supernova remnant, i.e from
few thousand kilometers to many tens of light years).
The physics of supernova explosions is complex and diverse
and in many parts, the explosion mechanism is
still an unsolved problem. Even the constraining parameters and ingredients
which makes the SN explode are still controversial (see e.g. the discussion in
\cite{Jan07}). It is possible that the identification of the
key aspects in the explosion may require seminal observations about the conditions in the supernova
core other than the 
indirect evidence such as explosion asymmetries, pulsar kicks or nucleosynthetic yields.
Such observations may involve the future
detection of strong neutrino signals and gravitational waves from
a galactic supernova in future. Detectable neutrino signals from a supernova was seen in the case
of SN 1987A, but since that target
was in a neighboring satellite galaxy (the Large Magellanic Cloud),
the number of neutrinos were too small to distinguish the fine points of theoretical issues.

Since nuclear astrophysics is not usually taught at the
master's level nuclear physics specialization
in our universities,
these lecture notes are meant to be an introduction to the subject and a
pointer to literature and Internet resources\footnote{
See for example, \cite{Hax99} for a course of nuclear astrophysics, and
the International Conference Proceedings under the title: ``Nuclei
in the Cosmos" for periodic research conferences in the field.
Valuable nuclear astrophysics datasets in machine readable
formats can be found at sites:\cite{Ornl99}, \cite{Lbl98}.
A new and updated version of the nuclear reactions rate library REACLIB
for astrophysics is now being maintained as a public, web-based version at
\cite{jina-reaclib08}. A complementary effort to develop software tools to streamline
the process of getting the latest and best information into this new library
is available at \cite{nucastrodata08} (see \cite{ms-smith08}).
Much of the material discussed in the first part of these notes can be found in textbooks
in the subject, see e.g. \cite{Rol88}, \cite{Cla68}, \cite{Ree68},
\cite{Bah89}, \cite{Arn96} etc. There is also a recent book on the subject by 
Richard Boyd \cite{Boy08} that among other topics describes terrestrial
and space born instruments 
operating in service to nuclear astrophysics. A Workshop on Solar
fusion cross sections for the pp chain and CNO cycle held in 2009
by the Institute of Nuclear Theory is expected to result in a Reviews
of Modern Physics article.
Supernovae of various types are
the sites where nuclear reactions in stars or explosions are of prime importance. For
Internet resources to two recent Schools on these topics, see
http://icts.tifr.res.in/sites/Sgrb/Programme and http://www.tifr.res.in/$\sim$sn2004.}. The
emphasis in the first part of these
lecture notes is on the nuclear reactions in the stars and how these
are calculated, rather than how stars evolve. The latter usually forms a core
area of stellar astrophysics. 

This article is organized essentially
in the same sequence that a massive star burns successively
higher atomic number elements in its core, until it collapses
and explodes in a supernova. The introductory part
discusses how the rates of thermonuclear reactions in (massive) stars are
calculated, what the different classes of reactions are and how the
stars (usually) manage to burn their fuels so slowly\footnote{These issues were discussed
in an earlier SERC School \cite{ray03}.}.
The middle part describes the nuclear physics during the collapse phase of the massive star.
The last part describes a few typical examples of what can be learned by optical, IR and X-ray
studies about nucleosynthesis and dynamics of explosion in supernovae and supernova remnants
such as Cassiopeia A, SN 1987A etc.
Only core-collapse supernovae are discussed in these lectures, those that arise
from massive stars (e.g. stars more massive than $8 M_{\odot}$ with typical solar metallicity
at the time they start burning hydrogen in their cores, i.e. at the ``Zero Age Main Sequence", 
before any mass was lost from their surface).
We shall not discuss the type Ia SNe\footnote{The supernovae are classified by astronomers
on the basis of their optical spectra at the time of their maximum light output. Those that
do not show the presence of hydrogen in their spectra are classified as type I SNe. A subclass of
them, type Ia's are believed to arise from thermonuclear explosions in the electron degenerate
cores of stars less massive than $8 M_{\odot}$ and are very useful to map the geometry of
our universe, because they serve as calibratable ``standard" candles. These ``thermonuclear
supernovae" are usually more luminous in the optical bands than the core-collapse varieties, but
while the former class put out several MeVs of energy
per nucleon, the core-collapse SNe or ccSNe, emit 
several {\it hundreds} of MeVs per nucleon from their {\it central engines} (mostly
in down-scattered neutrinos). Apart from the missing
hydrogen lines, the type Ia SNe show an
absorption ``trough" in their spectra around maximum light at 6150 \AA, due to blue shifted
Si II lines \cite{fil91a}. Of the other type I SNe which show no hydrogen and no
Si II trough, some show helium in their early
spectra (type Ib) and others have little or no helium and have a composition
enhanced in oxygen (type Ic) \cite{har90}.
These, (Ib and Ic) together with the type IIs constitute the core collapse SNe.} in these lectures.
Abundance of elements in our galaxy Milky Way 
give important information about how stars affect 
the element and isotopic evolution in various parts of the galaxy.
For a study of the evolution of
elements from C to Zn in the galactic halo and disk, with 
yields of massive stars and type Ia SNe, 
see \cite{gos00} and elsewhere in these Proceedings.

\section{Stars and their thermonuclear reactions}

While referring to Sir Ernest Rutherford ``breaking down the atoms
of oxygen and nitrogen",
Arthur Eddington remarked: ``what is possible in the Cavendish
Laboratory may not be too difficult in the sun" \cite{Edd20}.
Indeed this is the case, but for the fact that a star does
this by fusion reactions, rather than by transfer reactions, --
in the process giving out heat and light and manufacturing fresh elements.
Of all the light elements, hydrogen
is the most important one in this regard, because: a) it has a large
universal abundance, b) considerable energy evolution is possible
with it because of the large binding energies of nuclei that can
be generated from its burning and c) its small charge and mass
allows it to penetrate easily through the potential barriers of
other nuclei.
A star burns its fuel in thermonuclear reactions in the core
where the confinement of the fuel is
achieved in the star's own gravitational field. These reactions remain
``controlled", or self-regulated\footnote{There are however examples to the contrary
when thermonuclear reactions take place in an explosive manner, e.g.
when a whole white dwarf (resulting from an evolved intermediate mass star)
undergoes merger with another and explodes, as nuclear
fuel (carbon) is ignited under degenerate conditions,
such as in a type Ia supernova; explosive thermonuclear reactions also take place
in post-bounce core collapse supernovae, when the hydrodynamic shock ploughs through
unburnt Si- or O-layers in the mantle.}, as long as the stellar material
remains non-degenerate.

The recognition of the quantum mechanical tunneling effect prompted
Atkinson and Houtermans \cite{Atk29} to work out the qualitative treatment
of energy production in stars. They suggested 
that the nucleus serves as both a cooking pot and a trap.
Binding energy difference of four protons and two electrons
(the nuclear fuel) and their ash, the helium nucleus, some 26.7 MeV
is converted to heat and light that we receive from the sun\footnote{
Lord Kelvin surmised in the nineteenth century that the solar
luminosity is supplied by the gravitational contraction of the sun. Given
the solar luminosity, this immediately defined a solar lifetime (the so-called
Kelvin-Helmholtz time): $\tau_{KH} = G M_{\odot}^2/R_{\odot}L_{\odot} \sim
\rm few \times 10^7 \rm yr$. This turned out to be much shorter
than the estimated age
of the earth at that time known from fossil records, and led to a famous debate
between Lord Kelvin and the fossil geologist Cuvier. In fact, as noted above
modern estimates of earth's age are much longer
and therefore the need to maintain
sunshine for such a long time requires that the amount of energy
radiated by the sun during its lifetime is much larger than its gravitational
energy or its internal (thermal) energy:
$L_{\odot} \times t_{\rm life} \gg G M_{\odot}^2 / R_{\odot}$.
This puzzle was
resolved only later with the realization that the star can tap its
much larger nuclear energy reservoir in its core through thermonuclear
reactions. The luminosity of the sun however is determined by an
interplay of atomic and gravitational physics that controls the opacity,
chemical composition, the balance of pressure forces against gravity, etc.
Nuclear physics determines how fast nuclear reactions go under
under feedback control determined by the ambient conditions.
}.
The photons in the interior are scattered many a times, for tens of
millions of years, before they reach the surface of the sun.
Some of the reactions produce neutrinos, which because of their
small cross-section for interaction, are not stopped or scattered by
overlying matter, -- but stream out straight from the core.
Neutrinos are thus the best probes of the stellar core \cite{Dav68,hir87}, 
while the photons bear information from their surface of
last scattering -- the photosphere.

\subsection{Why do the stars burn slowly: a look at Gamow peaks}

The sun has been burning for at least 4.6 billion years\footnote
{Lord Rutherford \cite{Rut29} determined the age of a sample
of pitchblende, to be 700 million years,
by measuring the amount of uranium and radium and helium retained in the rock
and by calculating the annual output of alpha particles. The oldest
rock found is from Southwest Greenland: $\approx 3.8$ Gyr old
\cite{Bah89}. Radioactive dating of meteorites point to their formation and the
solidification of the earth about $4.55 \pm 0.07$ years ago \cite{Kir78}.
Since the sun and the solar system formed only slightly before,
their age at isolation and condensation
from the interstellar medium is taken to be 4.6 Gyr \cite{Fow77}.
}.
How does it manage to burn so slowly\footnote
{The Nobel prize citation of Hans Bethe (1967) who solved this problem, noted that this
`` concerns an old riddle. How has it been possible for the sun
to emit light and heat without exhausting its source not only
during the thousands of centuries the human race has existed
but also during the enormously long time when living beings
needing the sun for their nourishment have developed and
flourished on our earth thanks to this source? The solution of
this problem seemed even more hopeless when better
knowledge of the age of the earth was gained. None of the
energy sources known of old could come under consideration.
A very important part of his work resulted in eliminating a great
number of thinkable nuclear processes under the conditions at
the center of the sun, after which only two possible processes
remained..... (Bethe) attempted a thorough analysis of these and other
thinkable processes necessary to make it reasonably certain
that these processes, and only these, are responsible for the
energy generation in the sun and similar stars.
"}?
Under the ambient conditions in the core,
the relevant thermonuclear reaction cross sections are very small\footnote
{This makes the experimental verification of the
reaction cross-sections a
very challenging task, requiring in some cases, extremely
high purity targets and projectiles
so that relevant small event rates are not swamped by other reaction channels
and products (see Rolfs and Rodney, Chapter 5 \cite{Rol88}).}.
For reactions involving charged particles, nuclear
physicists often encounter cross-sections near the Coulomb barrier
of the order of millibarns. One can obtain a characteristic luminosity
$L_C$ based on this cross section and
the nuclear energy released per reaction \cite{Bah89} :

\begin{equation}
L_C \sim \epsilon N \Delta E / \tau_C
\end{equation}

where $\epsilon \approx 10^{-2}$ is the fraction of total number
of solar nuclei $N \sim 10^{57}$ that take part in nuclear fusion reactions
generating typically $\Delta E \sim 25$ MeV in hydrogen to helium conversion.
Here, the $\tau_C$ is the characteristic time scale for reactions,
which becomes minuscule 
for the cross-sections at the Coulomb barrier and the
ambient density and relative speed of the reactants etc:

\begin{equation}
\tau_C \sim {1\over n \sigma v} = {10^{-8} s \over [n/(10^{26} \rm \; cm^{-3})]
[\sigma / 1 \rm \; mbarn] [v /10^9 \rm \; cm \; s^{-1}] }
\end{equation}

This would imply a characteristic luminosity of
$L_c \approx 10^{20} L_{\odot}$,
even for a small fraction of the solar material taking part in the
reactions (i.e. $\epsilon \sim 10^{-2}$). If this was
really the appropriate cross-section for the reaction,
the sun would have been gone very quickly indeed.
Instead the cross-sections are much less than
that at the Coulomb barrier penetration energy (say at proton energies of
1 MeV), to allow for a long lifetime of the sun (in addition,
weak-interaction process gives a smaller cross-section for some reactions than
electromagnetic process, -- see Section 3.1).

\begin{figure}[htb]
\centering
\includegraphics[height=9cm]{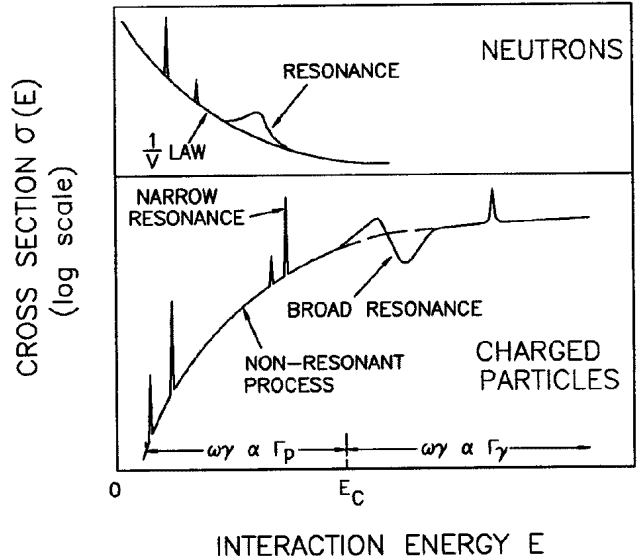}
\caption[]{
Dependence of total cross-sections on the interaction energy
for neutrons (top panel) and charged particles (bottom panel).
Note the presence
of resonances (narrow or broad) superimposed on a slowly varying
non-resonant cross-section (after \cite{Rol88}).
}
\label{fig: cross-sections}
\end{figure}

Stellar nuclear reactions can be either: a) charged particle reactions
(both target and projectile are nuclei) or b) neutral particle
(neutron) induced reactions. Both sets of reactions can go through
either a resonant state of an intermediate nucleus or can be a non-resonant
reaction. In the former reaction, the intermediate state could be a narrow
unstable state, which decays into other particles or nuclei. In general,
a given reaction can involve both types of reaction channels.
In charged particle induced reactions, the cross-section for both
reaction mechanisms drops rapidly with decreasing energy, due to
the effect of the Coulomb barrier (and thus it becomes more difficult
to measure stellar reaction cross-sections accurately).
In contrast, the neutron induced reaction cross-section is very large and
increases with decreasing energy (here, resonances may be superposed on a
smooth non-resonant yield which follows the
$1/v \sim 1/ \sqrt E$  dependence).
These reaction rates and cross-sections can be then directly
measured at stellar energies that are relevant
(if such nuclei are long lived or can be generated).
The schematic dependence of the cross-sections are shown in Fig.
\ref{fig: cross-sections}.
We shall not discuss the neutron capture elements in these notes. 

\subsection{Gamow peak and the astrophysical S-factor}

The sun and other ``main-sequence" stars (burning hydrogen
in their core quiescently) evolve very slowly by adjusting their central
temperature such that the average  thermal energy of a nucleus is small
compared to the Coulomb repulsion an ion-ion pair encounters. This is how
stars can live long for astronomically long times.
A central temperature $T \geq 10^7$K (or $T_7 \geq 1$, hereafter
a subscript x to a quantity, indicates that quantity in
units of $10^x$) is required for sufficient
kinetic energy of the reactants to overcome the Coulomb barrier
and for thermonuclear reactions involving
hydrogen to proceed at an effective rate, even though fusion reactions have
positive Q  values i.e. net energy is liberated out of the reactions.
The classical turning point radius for a projectile of charge $Z_2$ and
kinetic energy $E_p$ (in a Coulomb potential $V_C = Z_1 Z_2 e^2 / r $, and
effective height of the Coulomb barrier $E_C= Z_1 Z_2 e^2/ R_n = 550 \;\rm keV$
for a p + p reaction), is:
$r_{cl} = Z_1 Z_2 e^2 / E_p$.
Thus, classically a p + p reaction would proceed only when the kinetic energy
exceeds 550 keV. Since the number of particles traveling at a given speed
is given by the Maxwell Boltzmann (MB) distribution $\phi(E)$, only the tail
of the MB distribution above 550 keV is effective when
the typical thermal energy is 0.86 keV ( $T_9 = 0.01$). The ratio of
the tails of the MB distributions:
$\phi(550 \; \rm keV) / \phi(0.86 \; \rm keV)$
is quite minuscule, and thus classically at typical stellar temperatures
this reaction will be virtually absent.

\begin{figure}[htb]
\centering
\vspace*{-0.4cm}
\includegraphics[height=9cm]{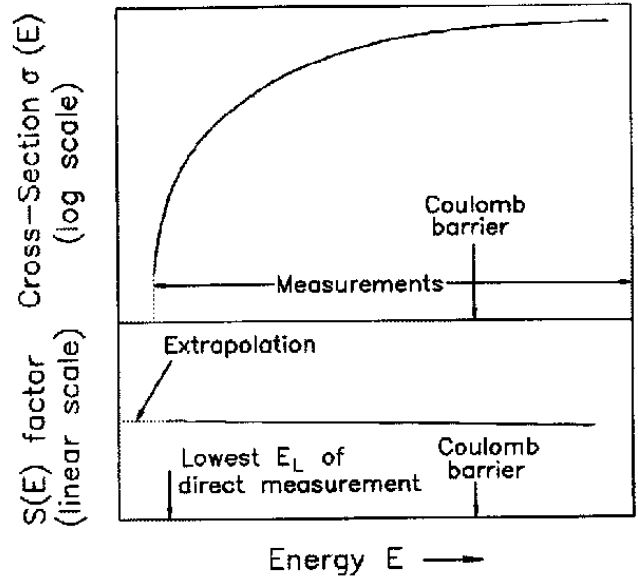}
\caption[]{
Cross-section and astrophysical S-factor for charged particle reactions
as a function of beam energy. The effective range of energy in stellar
interiors is usually far less than the Coulomb barrier energy $E_C$ or
the lower limit $E_L$ of laboratory measurements.
The y-scale is logarithmic for cross-section but linear for
S-factor; thus the cross section drops sharply in regions of
astrophysical interest, while the change is much less severe for the
S-factor. The extrapolation of laboratory data to lower
energies relevant for astrophysical situations is more reliable for 
S-factor.
}
\label{fig: SigmavsS-fac}
\end{figure}

Although classically a particle
with projectile energy $E_p$ cannot penetrate beyond the classical turning
point, quantum mechanically, one has a finite value of the squared
wave function at the nuclear radius $R_n : | \psi(R_n) |^2 $.
The probability that the incoming particle penetrates the barrier is:

\begin{equation}
 P = {| \psi(R_n) |^2 \over | \psi(R_c) |^2} 
\end{equation}

where $\psi(r)$ are the wave-functions at corresponding points. Bethe
\cite{Bet37} solved the Schroedinger equation for the Coulomb potential
and obtained the transmission probability:-

\begin{equation}
 P = \rm exp\bigg(-2 KR_c \big[{\rm tan^{-1}(R_c/R_n - 1)^{1/2}\over (R_c/R_n-1)^{1/2}}-{R_n\over R_c}\big]\bigg)
\end{equation}

with $K=[2\mu / \hbar^2(E_c - E)]^{1/2}$.
This probability reduces to a much simpler relation
at the low energy limit: $E \ll E_c$, which is equivalent to the classical
turning point $R_c$ being much larger than the nuclear radius $R_n$.
The probability is:

\begin{equation}
P = \rm exp(-2\pi\eta) = exp[-2\pi Z_1 Z_2 e^2/(\hbar v)] = exp[-31.3Z_1Z_2({\mu \over E})^{1/2}]
\end{equation}

where in the second equality, $\mu$ is the reduced mass in Atomic Mass Units
and E is the center of mass energy in keV. The exponential
quantity involving the square brackets in the
second expression is called the ``Gamow factor". The reaction cross-section
between particles of charge $Z_1$ and $Z_2$ has this exponential dependence
due to the Gamow factor. In addition, because the cross-sections are essentially
``areas": proportional to $\pi(\lambda/2\pi\hbar)^2 \propto 1/E$, it is
customary to write the cross-section, with these two energy dependences
filtered out:

\begin{equation}
\sigma(E) = {\rm exp(-2\pi\eta)\over E} S(E)
\end{equation}

where the factor $S(E)$ is called the astrophysical 
S-factor.
The S-factor may contain degeneracy factors due to spin, e.g.
$[(2J+1)/{(2J_1+1)(2J_2+1)}]$ as reaction cross-sections
are summed over final states and averaged over initial states.
Because the rapidly varying parts of the cross-section (with energy)
are thus filtered out, the S-factor is a slowly varying function
of center of mass energy, at least for the non-resonant reactions.
It is thus much safer to extrapolate $S(E)$ to the energies relevant
for astrophysical environments from the laboratory data, which is
usually generated at higher energies (due to difficulties of measuring
small cross-sections), than directly extrapolating the $\sigma(E)$,
which contains the Gamow transmission factor
(see Fig. \ref{fig: SigmavsS-fac}).
Additionally, in order to relate $\sigma(E)$ and $S(E)$, quantities
measured in the laboratory to these relevant quantities in the solar
interior, a correction factor $f_0$ due to the effects of electron screening
needs to be taken into account \cite{Sal54}.

In the stellar core with a temperature T, reacting particles
have many different velocities (energies) according to a Maxwell 
Boltzmann distribution:-

\begin{equation}
\phi(v) = 4\pi v^2 \bigg({\mu \over 2\pi kT}\bigg)^{3/2} \rm exp\bigg[-{\mu v^2 \over 2kT}\bigg]
 \propto E^{1/2} \; exp[-E/kT] 
\end{equation}

Nuclear cross-section or the reaction rates which also depend upon the
relative velocity (or equivalently the center of mass energy) therefore need
to be averaged over the thermal velocity (energy) distribution. Therefore,
the thermally averaged reaction rate per particle pair is:

\begin{equation}
<\sigma v> = \int_0^{\infty} \phi(v) \sigma(v) v dv
             = \big({8\over \pi \mu}\big)^{1/2}{1\over(kT)^{3/2}} \int_0^{\infty} \sigma(E) E \; \rm exp(-E/kT) dE 
\end{equation}

\begin{figure}[htb]
\centering
\hspace*{-0.5cm}
\includegraphics[height=7.5cm]{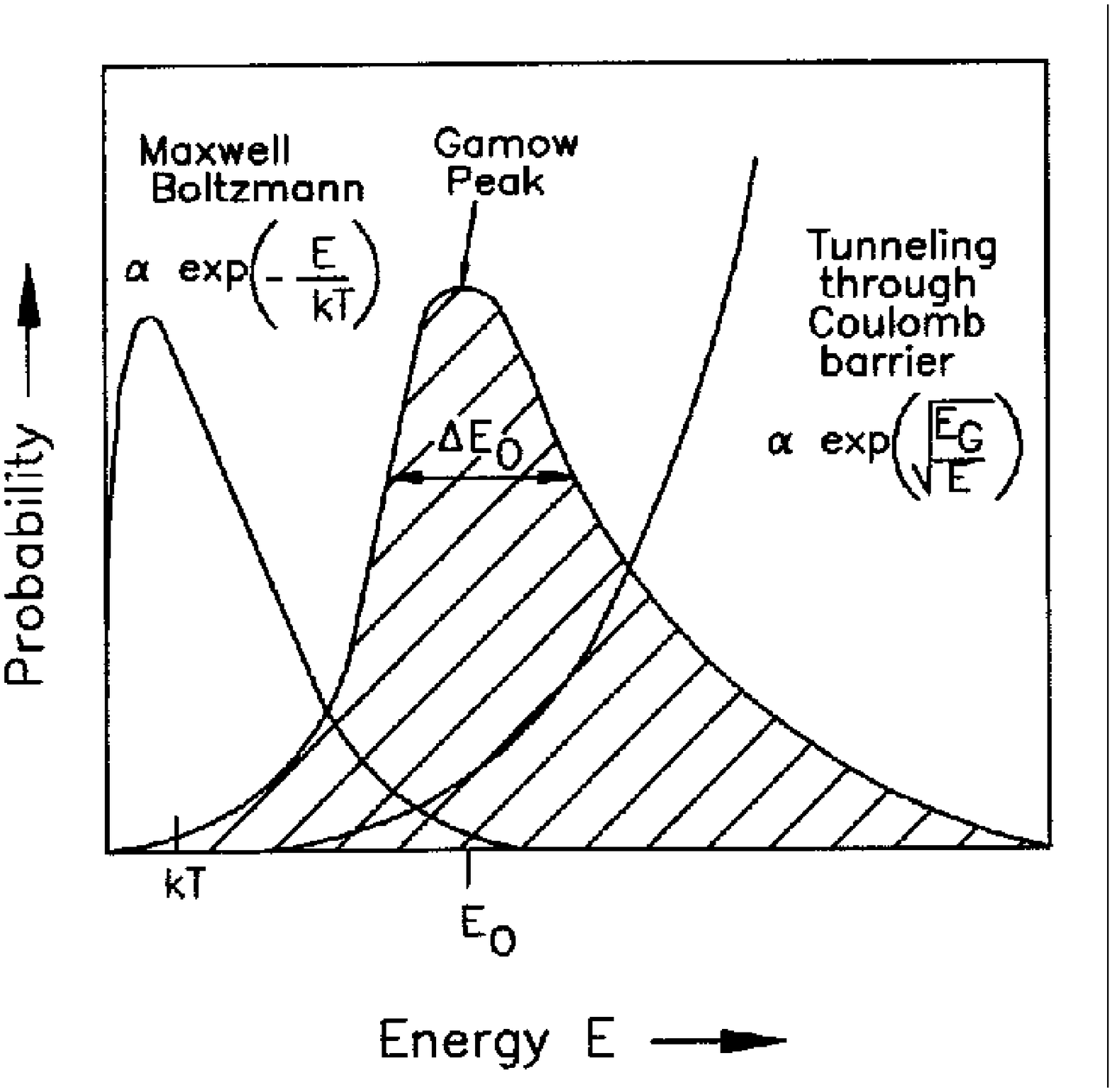}
\caption[]{
The Gamow peak is a convolution of the energy distribution of
the Maxwell Boltzmann probability and the quantum mechanical Coulomb
barrier transmission probability. The peak in the shaded region near
energy $E_0$ is the Gamow peak that gives the highest probability
for charged particle reactions to take place. Usually the Gamow peak is
at a much higher energy than $kT$, and in the figure the ordinate scale
(for the Gamow peak) is magnified with respect to those of the
M-B and barrier penetration factors. See also Table \ref{tab: parameters}.
}
\label{fig: gamowpeak}
\end{figure}

The thermally averaged reaction rate per pair is, utilizing the astrophysical
S-factor and the energy dependence of the Gamow-factor:

\begin{equation}
<\sigma v> = \big({8\over \pi\mu}\big)^{1/2}{1\over(kT)^{3/2}} \int_0^{\infty}\rm S(E) exp \big[ -{E\over kT} - {b\over \sqrt E} \big]dE 
\end{equation}

with $b^2 = E_G = 2\mu (\pi e^2 Z_1 Z_2/ \hbar)^2 = 0.978\mu Z_1^2Z_2^2$ MeV,
$E_G$ being called the Gamow energy.
Note that in the expression for the reaction rate above, at low energies,
the exponential term $\rm exp(-b/\sqrt E) = exp(-\sqrt(E_G/E))$ becomes
very small whereas at high energies the Maxwell Boltzmann factor
$\rm E^{1/2} \; exp(-E/kT)$ vanishes.
Hence there would be a peak (at energy, say, $E_0$)
of the integrand for the thermally averaged reaction rate per pair
(see Fig. \ref{fig: gamowpeak}).
The exponential part of the energy integrand can be approximated as:

\begin{equation}
 \rm exp \big[-{E\over kT} - b E^{-1/2}\big] \sim C \; exp \bigg[-\big({E-E_0\over \Delta/2}\big)^2\bigg]
\end{equation}

where
$$C= \rm exp(-E_0/kT - bE_0^{-1/2}) = exp(-3E_0/kT) = exp(-\tau)$$
$$E_0 = (b kT/2)^{{2\over3}} = 1.22 \rm keV (Z_1^2 Z_2^2 \mu T_6^2)^{{1\over3}}$$

$$\Delta = 4(E_0 kT/3)^{{1\over2}} = 0.75 \rm keV (Z_1^2 Z_2^2 A T_6^5)^{{1\over6}}$$

Since most stellar reactions happen in a fairly narrow band of energies,
S(E) will have a nearly constant value over this band averaging to
$S_0$. With this, the reaction rate per pair of particles, turns out to be:
\begin{equation}
<\sigma v> = \big[{2\over kT}\big]^{3\over2} {S_0\over\pi\mu} \int^{\infty}_0e^{-\tau -4({E-E_0\over \Delta})^2}dE = 4.5 10^{14} {S_0\over AZ_1Z_2} \tau^2 e^{-\tau} \rm cm^3s^{-1}
\end{equation}
Here,

\begin{equation}
\tau = 3 E_0 /kT = 42.5 (Z_1^2 Z_2^2 \mu /T_6)^{{1\over3}}
\end{equation}

The maximum value
of the integrand in the above equation is:
$$ \rm I_{max} = exp(-\tau)$$
The values
of $\rm E_0, I_{max}, \Delta,$ etc., apart from the Coulomb barrier for
several reactions are tabulated in Table \ref{tab: parameters}
for $T_6 = 15$.

As the
nuclear charge increases, the Coulomb barrier increases, and the Gamow peak
$E_0$ also shifts towards higher energies. Note how rapidly the maximum
of the integrand $I_{max}$ decreases with the nuclear charge and the Coulomb
barriers. The effective width $\Delta$ is a geometric mean of $E_0$ and
kT, and $\Delta/2$ is much less rapidly varying between reactions (for
$kT \ll E_0$). The rapid variation of $I_{max}$  indicates that of several
nuclei present in the stellar core, those nuclear pairs will have the largest
reaction rates, which have the smallest Coulomb barrier.
The relevant nuclei will be consumed most rapidly at that stage.
(Note however
that for the p+p reaction, apart from the Coulomb barrier, the strength of
the weak force, which transforms a proton to a neutron also
comes into play).

When nuclei of the smallest Coulomb barrier are
consumed, there is a temporary dip in the nuclear generation rate, and the
star contracts gravitationally until the temperature rises to a point where
nuclei with the next lowest Coulomb barrier will start burning. At that
stage, further contraction is halted.
The star goes through well defined stages
of different nuclear burning phases in its core dictated
by the height of the Coulomb barriers of the fuels.
Note also from the Table \ref{tab: parameters},
how far $E_0$, the effective mean energy
of reaction is below the Coulomb barrier at the relevant temperature.
The stellar burning is so slow because the reactions are taking place at
such a far sub-Coulomb region, and this is why the stars can last so long.
\begin{table}[hb]
\vspace*{-12pt}
\caption[]{
Parameters of the thermally averaged reaction rates at $T_6 =15$.
}
\begin{center}
\begin{tabular}{llllll}
\hline\\[-10pt]
Reaction & Coulomb & Gamow & $I_{max}$  & {\phantom{$00$}}$\Delta$ & {\phantom{$0$}}$(\Delta)I_{max}$\\
         & Barrier & Peak ($E_0$) & ($e^{-3E_0/kT}$)    &           & \\
         & (MeV)   & (keV)   &                       & {\phantom{$0$}}(keV)      &\\
\hline\\[-10pt]
p + p     & 0.55& 5.9 & $1.1 \times 10^{-6}$ &
{\phantom{$00$}}6.4 & $7 \times 10^{-6}$\\
p + N     & 2.27& 26.5 & $1.8 \times 10^{-27}$ & {\phantom{$0$}}13.6 & $2.5 \times 10^{-26}$\\
$\alpha$ + C$^{12}$ & 3.43& 56 & $3 \times 10^{-57}$ & {\phantom{$0$}}19.4 & $5.9 \times 10^{-56}$\\
O$^{16}$ + O$^{16}$ & 14.07& 237 & $6.2 \times 10^{-239}$ &
{\phantom{$0$}}40.4 & $2.5 \times 10^{-237}$\\
\hline
\end{tabular}
\end{center}
\label{tab: parameters}
\end{table}

The above discussion assumes that a bare nuclear Coulomb potential
is seen by the charged projectile. For nuclear reactions measured
in the laboratory, the target nuclei are in the form of atoms with
electron cloud surrounding the nucleus and giving rise to a screened
potential -- the total potential then goes to zero outside the atomic radius.
The effect of the screening is to reduce the effective height of the
Coulomb barrier.
Atoms in the stellar interiors are in most cases in highly stripped state,
and nuclei are immersed in a sea of free electrons which tend to cluster
near the nucleus. When the stellar density increases, the so called
Debye-Huckel radius $R_D = (kT/ 4\pi e^2 \rho N_A \xi)^{1/2}$ ,
(here: $\xi = \sum_i (Z^2_i + Z_i) X_i/A_i$) which is
a measure of this cluster ``radius", decreases, and the effect of shielding
upon the reaction cross-section becomes more important.
This shielding effect enhances
thermonuclear reactions inside the star. The enhancement factor
$f_0 = \rm exp (0.188 Z_1 Z_2 \xi \rho^{1/2} T_6^{-3/2}$,
varies between 1 and 2 for typical densities and compositions \cite{Sal54}
but can be large at high densities.

\section{Hydrogen burning: the pp chain}

The quantitative aspects of the problem of solar
energy production with details of known nuclear physics of converting
hydrogen into helium
was first worked out by von Weizs\"acker (1937-38) \cite{Wei37}, \cite{Wei38}
and Bethe \& Critchfield (1938-1939) \cite{Bet38},
which made it clear that two different sets of
reactions : the p-p chains and the CN cycle can do this conversion.
This happens in the core of the star initially (at the ``main sequence"
stage), and then later in the
life of a star in a shell of burning hydrogen around an inert core of He.

In the first generation of stars in the galaxy only the p-p cycle may have
operated. In second generation, heavier elements like C, N from the ashes
of burning in previous stars are available and they too can act as
catalysts to have thermonuclear fusion of hydrogen to helium.
Since in the very first generation, the heavier elements like C, N were practically
absent, all stars including the massive stars, burnt hydrogen through the p-p cycle.
[A recent discovery (\cite{Chr02}) of a low-mass star with an
iron abundance as
low as 1/200,000 of the solar value (compare the previous record
of lowest iron abundance less than 1/10,000 that of the sun), suggests
that such first generation stars are still around].

\begin{figure}[htb]
\centering
\hspace*{-0.6cm}
\vspace*{-0.5cm}
\includegraphics[height=9.5cm,width=12.2cm,angle=-0.5]{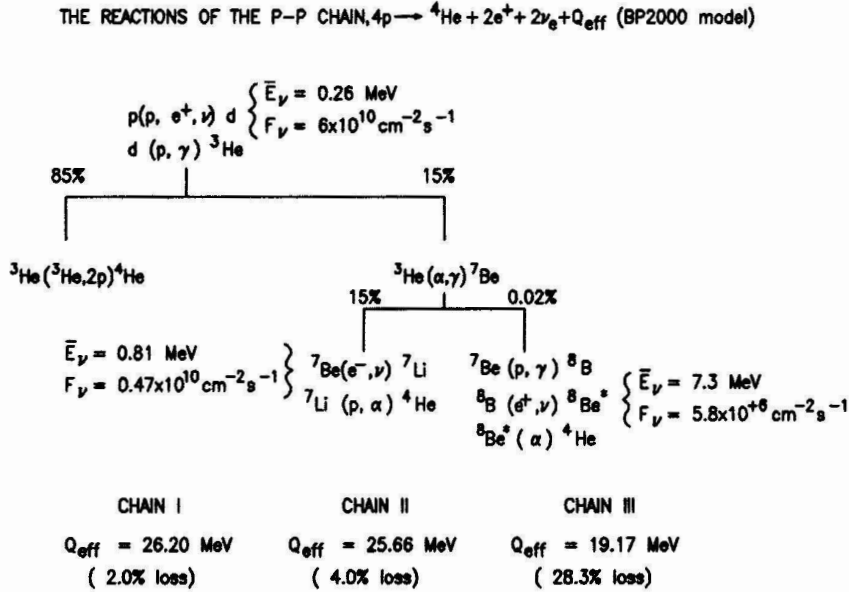}
\caption[]{
The p-p chain starts with the formation of deuterium
and $^3He$. Thereafter,  $^3He$ is consumed in the sun 85\% of
the time through ppI chain, whereas ppII and ppIII chains
together account for 15\% of the time in the Bahcall Pinsonneault 2000
solar model. The ppIII chain occurs only $0.02 \%$ of the time,
but the $^8B$ $\beta^+$-decay provides the higher energy neutrinos
(average $\bar E_{\nu} = 7.3 \; \rm MeV$). The net result of the chains
is the conversion of four protons to a helium, with the effective
Q-values (reduced from 26.73 MeV) as shown, due to loss of energy
in escaping neutrinos. See \cite{Hax08,Gar08} for updated branching ratios
and neutrino fluxes for BPS2008(AGS) model.
}
\label{fig: ppchain}
\end{figure}

The sun with a central temperature of 15.7 million degrees, ($T^c_{6\odot}
= 15.7$) burns by p-p chains. Slightly more massive star (with
central temperature $T_6  \geq 20$) burns H by the CNO cycle also.
Davis et al.s' solar neutrino experiment \cite{Dav68}, which in 1968 had only
an upper limit of the neutrino flux, itself put a limit of
less than 9\% of the sun's energy is produced by the carbon-nitrogen cycle
(the more recent upper limit \cite{Bah03} is $7.3\%$, from an analysis
of several solar neutrino experiments, including the Kamland
measurements). Note however that
for the standard solar model, the actual contribution of CNO cycle
to solar luminosity is $\sim 1.5 \%$ \cite{Bah89}).
In CNO cycle, nuclei such as C, N, O serve as ``catalysts" do in a
chemical reaction.
The pp-chain and the CNO cycle reaction sequences are illustrated
in Figs. \ref{fig: ppchain} and \ref{fig: CNOcycle}.

The pp-chain begins with the reaction $\rm p + p \rightarrow d + e^+ + \nu_e$.
Bethe and Critchfield \cite{Bet38} showed that weak nuclear reaction
is capable of converting a proton into a neutron during the brief
encounter of a scattering event. (This reaction overcomes the impasse
posed by the instability of $^2$He in the $\rm p + p \rightarrow  ^2He $ reaction
via the strong or electromagnetic interactions, and as the next nucleus
$^3$Li reachable via these interactions is also unstable as a final product).
Since a hydrogen atom is less massive
than a neutron, such a conversion would be endothermic
(requiring energy), except for the fact that a neutron in a deuterium nucleus
$^2$D can form a bound state with the proton with a binding energy of
2.224 MeV -- thus making the reaction exothermic with an available kinetic
energy of 0.42 MeV. The released positron soon pair annihilates into 
photons making the total energy released to be 1.442 MeV.

Because of the low Coulomb barrier, in the p+p reaction ($E_c = 0.55$ MeV),
a star like the sun would have consumed all its hydrogen quickly
(note the relatively large value of $(\Delta) I_{max}$ in
Table \ref{tab: parameters}),
were it not slowed down by the weakness of the weak interactions.
The calculation of probability of deuteron formation consists of two
separate considerations: 1) the penetration of a mutual potential
barrier in a collision of two protons in a thermal bath and 2) the
probability of the $\beta$-decay and positron and neutrino emission.
Bethe and Critchfield used the original Fermi theory (point interaction)
for the second part, which is adequate for the low energy process.

\subsection{Cross-section for deuterium formation}

The total Hamiltonian $H$ for the p-p interaction can be written as a
sum of nuclear term $H_n$ and a weak-interaction term $H_w$. As the weak
interaction term is small compared to the nuclear term, first order
perturbation theory can be applied and Fermi's ``Golden rule",
gives the differential cross-section as:

\begin{equation}
 d\sigma = {2\pi \rho(E) \over \hbar v_i} |<f | H_w | i>|^2 
\end{equation}

here $\rho (E) = dN/dE$, is the density of final states in the interval $dE$
and $v_i$ is the relative velocity of the incoming particles.
For a given volume V, the number of states dn between p and p+dp is:-

\begin{equation}
 dN = dn_e dn_{\nu} = (V {4 \pi p_e^2 dp_e \over h^3}) (V {4 \pi p_{\nu}^2 dp_{\nu} \over h^3}) 
\end{equation}

By neglecting the recoil energy of deuterium (since this is much heavier
than the outgoing positron in the final state) and neglecting the mass of the
electron neutrino, we have:

\begin{equation}
E=E_e + E_{\nu}=E_e + c p_{\nu}
\end{equation}

and $dE=dE_{\nu} = c p_{\nu}$, for a given $E_e$ and,

\begin{equation}
 \rho(E) = dN(E)/dE = dn_e (dn_{\nu}/dE) = {16 \pi^2 V^2 \over c^3 h^6} p_e^2 (E-E_e)^2 dp_e = \rho(E_e) dp_e 
\end{equation}

The matrix element that appears in the differential cross section, may be
written in terms of the initial state wave function $\Psi_i$ of the two protons
in the entrance channel and the final state wave function
$\Psi_f$ as:

\begin{equation}
H_{if} = \int [\Psi_d \Psi_e \Psi_{\nu}]^* H_{\beta} \Psi_i d\tau 
\end{equation}

If the energy of the electron is large compared to $Z\times \rm Rydberg$
(Rydberg $R_{\infty}= 2 \pi^2 m e^4 /ch^3$), then a plane wave approximation
is a good one:
$\Psi_e = 1/(\sqrt V)\rm exp(i {\bf \vec{k_e}} . {\bf \vec{r}})$
where the wave-function is normalized over volume V.
(For lower energies, typically 200 keV or less, the electron wave-function
could be strongly affected by nuclear charge (see \cite{Sch83})).
Apart from this, the final state wave function: $[\Psi_d \Psi_e \Psi_{\nu}]$
has a deuteron part $\Psi_d$ whose radial part rapidly vanishes
outside the nuclear domain ($R_0$), so that the integration need not
extend much beyond $r \simeq R_0$ (for example, the deuteron
radius $R_d = 1.7$ fm). Note that because of the Q-value
of 0.42 MeV for the reaction, the kinetic energy of the electron
($K_e \leq 0.42$ MeV) and the average energy of the neutrinos
($\bar E_{\nu} = 0.26$ MeV) are low enough so that for both
electrons and neutrino wave-functions, the product
$k R_0 \leq 2.2\times 10^{-3}$
and the exponential can be approximated by the first term of the
Taylor expansion:

\begin{equation}
\Psi_e= 1/(\sqrt V)\rm [1+i({\bf \vec{k_e}}.{\bf \vec{r}})] \sim 1/(\sqrt V)
\end{equation}
and
$$\Psi_{\nu} \sim 1/(\sqrt V)$$
Then the expectation value of the Hamiltonian, for a 
coupling constant $g$ is:

\begin{equation}
H_{if} = \rm \int [\Psi_d \Psi_e \Psi_{\nu}]^* H_{\beta} \Psi_i d\tau = {g\over V} \rm \int [\Psi_d ]^* \Psi_i d\tau
\end{equation}

The $d\tau$ integration can be broken into space and spin parts $M_{space}$ and
$M_{spin}$: 

\begin{equation}
d\sigma = {2\pi \over \hbar v_i} {16 \pi^2 \over c^3 h^6} g^2 M^2_{spin} M^2_{space} p_e^2 (E-E_e)^2 d p_e 
\end{equation}

The total cross-section upto an electron energy of $E$ 
is proportional to:

\begin{equation}
\int_0^E p_e^2 (E-E_e)^2 d p_e = {(m_e c^2)^5 \over c^3} \int_1^W (W_e^2 -1)^{1/2} (W-W_e)^2 W_e d W_e 
\end{equation}

where $W= (E+m_ec^2)/m_ec^2$.
The integral over W can be shown as:

\begin{equation}
f(W)=(W^2-1)^{1/2}[{W^4\over30}-{3W^2\over20}-{2\over15}]+{W\over4}\; \rm ln[W+(W^2-1)^{1/2}]
\end{equation}

so that:
\begin{equation}
\sigma = { m_e^5 c^4 \over 2 \pi^3 \hbar^7} f(W) g^2 M_{space}^2 M_{spin}^2
\end{equation}

At large energies, the factor $f(W)$ behaves as:

\begin{equation}
f(W) \propto W^5 \propto {1\over 30} E^5
\end{equation}

The final state nucleus (deuterium in its ground state)
in the reaction: $\rm p + p \rightarrow d + e^+ + \nu_e$, has
$J_f^{\pi} = 1^+$, with a predominant relative orbital angular momentum
$l_f = 0$ and $S_f = 1$ (triplet S-state). For a maximally probable
super-allowed transition, there is no change
in the {\it orbital} angular momentum between the initial and final states
of the nuclei. Hence for such transitions, the initial state two
protons in the $\rm p+p$ reaction
must have $l_i = 0$. Since the protons are identical particles,
Pauli principle requires $S_i = 0$, so that the total wave-function will be
antisymmetric in space and spin coordinates. Thus, we have a process:
$ | S_i =0, l_i = 0 >  \rightarrow  | S_f =1, l_f = 0 > $.
This is a pure Gamow-Teller\footnote{
In the beta-decay {\it allowed} approximation, we neglect the
variation of the lepton wave-functions
over the nuclear volume and the nuclear momentum 
(this is equivalent to neglecting all total lepton
orbital angular momenta $L > 0$). The total angular momentum
carried off by the leptons is their total spin: i.e. $S=1$ or 0,
since each lepton has ${s=1\over2}$. When the lepton spins in the
final state are anti-parallel, $s_e + s_{\nu} = s_{tot} = 0$ the
process is the Fermi transition with Vector coupling constant
$g = C_V$ (e.g. a pure Fermi decay:
$^{14}O (J_i^{\pi} = 0^+) \rightarrow ^{14}N(J^{\pi}_f = 0^+)$).
When the final state lepton spins are parallel,
$s_e + s_{\nu} = s_{tot} =1$, the process is Gamow-Teller with $g=C_A$.
For Fermi coupling, there is no change in the (total) angular
momentum between the initial and final states of the nuclei ($\Delta J=0$).
For the Gamow-Teller coupling, the selection rules are: $\Delta J =0$
or $\pm 1$ (but the possibility $\Delta J = 0$ is excluded 
between two states of zero angular momentum).
The size of the matrix element for a transition depends
on the overlap of the wave-functions in the initial and final states.
In the case of ``mirror pair" of nuclei (the nucleus $A_Z = (2Z+1)_Z$
is the mirror of the nucleus $(2Z+1)_{Z+1}$), the wave-functions are
very much alike as shown through simple heuristic arguments
\cite{Fer51}. For these nuclei, $ft$-values range from
$\sim 1000 - 5000$ and are called super-allowed
transitions. For super-allowed transitions, which have maximum
decay probabilities, there are no changes in the {\it orbital} angular
momentum between the initial and final states of the nuclei.
In the $\rm p \; + \; p \rightarrow D \; + e^+ + \; \nu_e$ reaction,
the initial proton state is antisymmetric to an interchange of
space and spin coordinates and the final deuteron is symmetric in
this respect. (In fact when the two protons are in the S state (which
is most favorable for their close approach), their spins will be
anti-parallel (a singlet state) whereas the ground state of the deuteron
is a triplet S state). If this were the complete description of the
exchange symmetry properties of the Gamow-Teller transition
(permitting a change of spin direction of the proton as it transforms
to a neutron, changing the total spin by one unit) advocated here
this would actually be forbidden. However in the use of configuration
space in beta-decay process one must include isotopic spin as well.
The $^1S$ state of the two protons is symmetric to exchange of this
coordinate, whereas the deuteron (consisting of both a proton and a
neutron) function is antisymmetric in this coordinate. In the complete
coordinate system the transition is from an initial antisymmetric state to
another antisymmetric final state accompanied by a positron emission
(\cite{BetCri38}).
} 
transition with coupling constant
$g= C_A$ (the axial vector coupling component can be obtained,
from the pure GT decay $^6He(0^+) \rightarrow ^6Li(1^+)$).

\begin{figure}[htb]
\centering
\includegraphics[height=12cm,width=11cm]{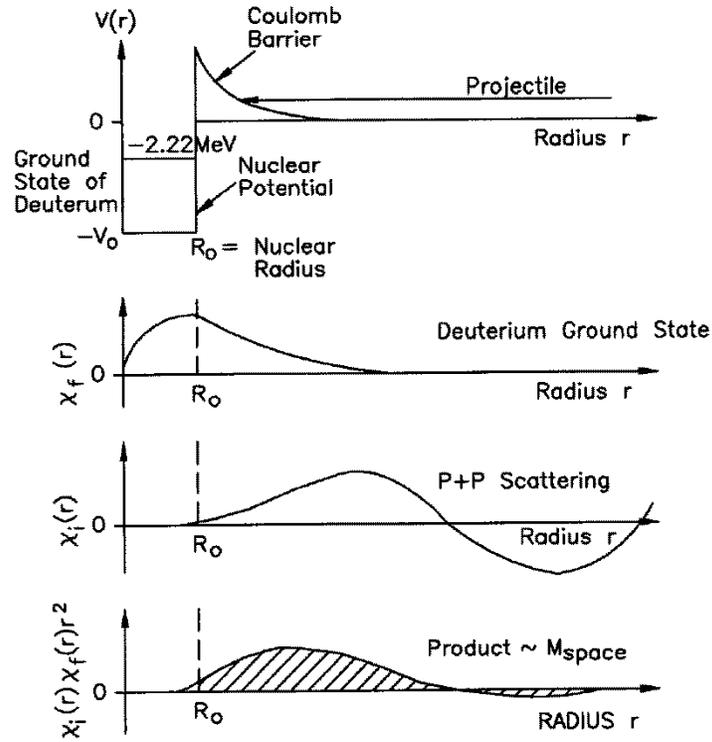}
\vspace*{-0.7cm}
\caption[]{
Schematic representation (after \cite{Rol88}) of the numerical calculation
of the spatial part of the matrix element $M_{space}$ in the
$\rm p + p \rightarrow \; d + e^+ + \nu_e$ reaction. The top part
shows the potential well of depth $V_0$ and nuclear radius $R_0$ of
deuterium with binding energy of $-2.22 \rm \; MeV$. The next
part shows the radius dependence of the deuterium radial wave
function $\chi_d(r)$. The wave-function extends far outside the nuclear
radius with appreciable amplitude due to the loose binding of deuterium
ground state. The p-p wave-function $\chi_{pp}(r)$
which comprise the $l_i = 0$ initial
state has small amplitude inside the final nuclear radius.
The radial part of the integrand entering into the calculation of
$M_{space}$ is a convolution of both $\chi_d$ and $\chi_{pp}$
in the second and third panels and is given with the hatched shading
in the bottom panel.
It has the major contribution far outside the nuclear radius.
}
\label{fig: overlapintegral}
\end{figure}

The spin matrix element in the above expression for energy integrated
cross-section $\sigma$, is obtained from summing over the final states
and averaging over the initial states {\it and} dividing by 2 to take
into account that we have two identical particles in the initial state.
Thus,

\begin{equation}
\lambda = {1\over \tau} = {m^5 c^4 \over 2 \pi^3 \hbar^7 v_i} f(W) g^2 {M^2_{space} M^2_{spin} \over 2}
\end{equation}

where, $M^2_{spin} ={(2J+1)\over(2J_1+1)(2J_2+1)} =3$. And the space matrix
element is:

\begin{equation}
M_{space} = \int_0^{\infty} \chi_f(r) \chi_i(r) r^2 dr
\end{equation}

in units of $\rm cm^{3/2}$.
The above integral contains the radial parts of the nuclear
wave-functions $\chi(r)$, and involves Coulomb wave-functions for barrier
penetration at (low) stellar energies. The integral
has been evaluated by
numerical methods (\cite{Fri51}), and Fig. \ref{fig: overlapintegral}
shows schematically
how the $M_{space}$ is evaluated for the overlap of
the deuterium ground state wave-function with the initial
pair of protons state. (See also
\cite{Sal52}, \cite{Bah69}
for details of calculations
of the overlap integral and writing the astrophysical S-factor
in terms the beta decay rate of the neutron
\cite{Sal52} which takes into account of radiative corrections
to the axial-vector part of the neutron decay through an effective
matrix element, the assumption being that these are the same
as that for the proton beta decay in the pp reaction above).
In the overlap integral
one needs only the S-wave part
for the wave-function of the deuteron $\psi_d$,
as the D-wave part makes
no contribution to the matrix element \cite{Fri51}, although its contribution
to the normalization has to be accounted for. The wave-function
of the initial two-proton system $\psi_p$ is normalized to a plane
wave of unit amplitude, and again only the S-wave part is needed.
The asymptotic form of $\psi_p$
(well outside the range of nuclear forces)
is given in terms of regular and
irregular Coulomb functions and has to be defined through quantities
related to the S-wave phase shifts in p-p scattering data).
The result is a minuscule total cross-section of $\sigma = 10^{-47} \rm cm^2$
at a laboratory beam energy of $E_p = 1 \; \rm MeV$, which cannot
be measured experimentally even with milliampere beam currents.

The reaction $\rm p + p \rightarrow d + e^+ + \nu_e$ is a non-resonant
reaction and at all energies the rate varies smoothly with
energy (and with stellar temperatures), with $S(0)=3.8\times 10^{-22}
\rm \; keV \; barn$ and $dS(0)/dE = 4.2\times10^{-24} \rm \; barn$.
At for example, the central temperature of the sun $T_6 =15$, this gives:
$<\sigma v>_{pp} = 1.2\times10^{-43} \rm\; cm^3 \; s^{-1}$.
For density in the center of the sun $\rho = 100 \; \rm gm\; cm^{-3}$
and and equal mixture of hydrogen and helium ($X_H = X_{He} =0.5$),
the mean life of a hydrogen nucleus against conversion to deuterium
is $\tau_H(H) = 1/N_H <\sigma v>_{pp} \sim 10^{10} \rm yr$. This is
comparable to the age of the old stars. The reaction is so slow
primarily because of weak interactions and to a lesser extent due
to the smallness of the Coulomb barrier penetration factor
(which contributes a factor $\sim 10^{-2}$ in the rate), and is
the primary reason why stars consume their nuclear fuel of hydrogen
so slowly. For a calculation of the weak capture of protons on protons
using calculated wavefunctions obtained from modern, realistic high
precision interactions, see \cite{Sch98}

\subsection{Deuterium burning}

Once deuterium is produced in the weak interaction mediated $\rm p + p$
reaction, the main way this is burnt in the
sun turns out to be:

\begin{equation}
d + p \rightarrow \; ^3He \; + \gamma 
\end{equation}

\begin{figure}[htb]
\centering
\hspace*{-1.3cm}
\includegraphics[height=9cm]{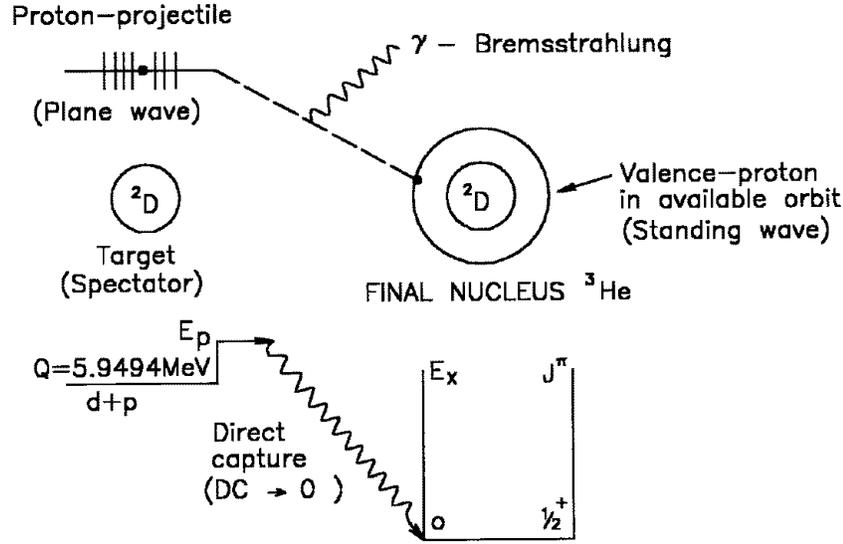}
\vspace*{-0.7cm}
\caption[]{
The Direct Capture reaction $d(p, \gamma)^3He$  to form $^3He$
in its ground state. The proton projectile (shown as a plane wave)
radiates away a bremsstrahlung photon to be captured in a ``valence"
orbital around the $^2D$.
}
\label{fig: directcapture}
\end{figure}

This is a non-resonant direct capture reaction to
the $^3He$ ground state with a Q-value of 5.497 MeV and
$S(0) = 2.5 \times 10^{-3} \rm keV \; barn$.
The angle averaged cross-sections measured as a function
of proton + deuterium center of mass energy, where the capture
transitions were observed in gamma-ray detectors at several
angles to the incident proton beam direction, are well explained
by the direct capture model (see Fig. \ref{fig: directcapture} after
\cite{Rol88}). The LUNA collaboration \cite{LUN02}
has measured the cross section
down to a 2.5 keV c.m. energy, well below the solar Gamow peak using a
segmented Bismuth germanate (BGO) gamma-ray detector and found the
S(E) factor to be in fair agreement with extrapolation of data at higher
energies. 

The reactions comprising the rest of
the (three) pp-chains start out with the predominant product of
deuterium burning: $^3He$ (manufactured from $d + p$ reaction)
as the starting point.
The only other reactions with a $S(0)$ greater than the above
are:
$\rm d(d,p)t \;, d(d,n)^3He \;, d(^3He,p)\\^4He$, and
$d(^3He, \gamma)^5Li$. However, because of the overwhelmingly
large number of protons in the stellar thermonuclear reactors,
the process involving protons on deuterium dominates. The rate
of this reaction is so fast compared to its precursor:
$p+p \rightarrow d + e^+ \nu_e$, that the overall rate
of the pp-chain is not determined by this reaction.

One can show that the abundance ratio of deuterium to hydrogen
in a quasi-equilibrium has an extremely small value, signifying
that deuterium is destroyed in thermonuclear burning. The time
dependence of deuterium abundance D is:

\begin{equation}
{d D \over dt} = r_{pp} - r_{pd} = {H^2\over2} <\sigma v>_{pp} - HD <\sigma v>_{pd} 
\end{equation}

The self regulating system eventually reaches a state of quasi-equilibrium
and has:

\begin{equation}
(D/H) = <\sigma v>_{pp} / (2<\sigma v>_{pd}) = 5.6\times 10^{-18}
\end{equation}

at $T_6 =5$ and $1.7\times 10^{-18}$ at $T_6 = 40$. For the solar system
however, this ratio is $1.5\times10^{-4}$ and the observed $(D/H)_{obs}$
ratio in the cosmos is $\sim 10^{-5}$. The higher cosmic ratio is due
to primordial nucleosynthesis in the early phase of the universe before
the stars formed. (The primordial deuterium abundance
is a key quantity used to determine the baryon density
in the universe). Stars only destroy the deuterium in their core
due to the above reaction.

\subsection{$\rm ^3He$ burning}

The pp-chain-I is completed (see Fig. \ref{fig: ppchain})
through the burning of $^3He$ via the reaction:

\begin{equation}
\rm ^3He \; + \; ^3He \rightarrow \;p \; + \; p \; + \; ^4He
\end{equation}

with an S-factor: $S(0) = 5500 \; \rm \; keV \; barn$ and Q-value = 12.86 MeV.
In addition, the reaction:

\begin{equation}
\rm ^3He \; + \; D \rightarrow \; ^4He \; + p
\end{equation}

has an S-factor: $S(0) = 6240 \; \rm keV \; barn$, but since the
deuterium concentration is very small as argued above,
the first reaction dominates the destruction of $^3He$
even though both reactions have comparable $S(0)$ factors.
The cross section for the former reaction near the Gamow energy
for the sun, has been measured in \cite{Ita03}.

$^3He$ can also be consumed by reactions with $^4He$ (the latter
is pre-existing from the gas cloud from which the star formed
and is synthesized in the early universe and in Pop III objects).
These reactions proceed
through Direct Captures and lead to the ppII and ppIII parts
of the chain (happening $15\%$ of the time). Note
that the reaction $^3He(\alpha, \gamma)^7Be$ together with the
subsequent reaction: $^7Be(p, \gamma)^8B$ control the production
of high energy neutrinos in the sun and are particularly important
for the $^{37}Cl$ solar neutrino detector constructed by Ray Davis and
collaborators.

\subsection{Reactions involving $^7Be$}

As shown in Fig. \ref{fig: ppchain}, about 15\% of the time, $^3He$ is
burned with $^4He$ radiatively to $^7Be$. Subsequent reactions
involving $^7Be$ as a first step in alternate ways complete the fusion
process: $4 H \rightarrow  \; ^4He$ in the ppII and ppIII chains.

\subsubsection{Electron capture process}

\begin{figure}[htb]
\centering
\includegraphics[height=6.0cm,width=9.7cm, angle=-0.0]{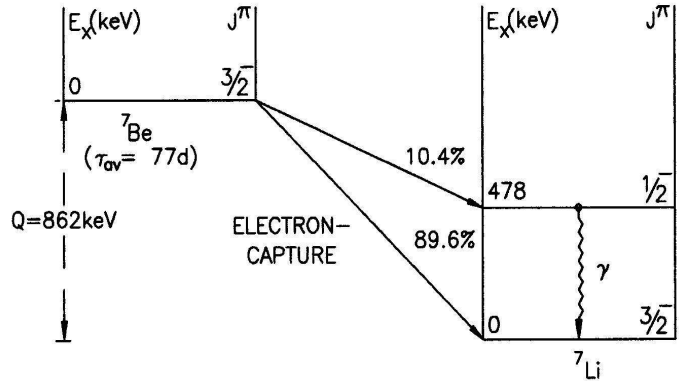}
\caption[]{
Electron capture on $^7Be$ nucleus. The capture proceeds 10.4\% of the time
to the first excited state of $^7Li$ at 478 keV, followed by a decay
to the ground state by the emission of a photon. The average energy of
the escaping neutrinos (which are from the ppII chain) is 814 keV.
}
\label{fig: ecapture7Be}
\end{figure}

The first step of the ppII chain is the electron capture reaction
on $^7Be$ : $^7Be \; + \; e^- \rightarrow \; ^7Li \; + \; \nu_e$
(see Fig \ref{fig: ecapture7Be}).
This decay goes both to the ground state of $^7Li$ as well as to its
first excited state at $E_X = 0.478 \; \rm keV, \; J^{\pi}={1\over2}^-)$
 -- the percentage of decays to the excited state being 10.4 \% in the
laboratory. The energy
released in the reaction with a Q-value of $0.862 \; \rm keV$
is carried away by escaping mono-energetic neutrinos with
energies: $E_{\nu} = 862$ and 384 keV. The measured laboratory mean life
of the decay is $\tau = 76.9 \rm d$.
The capture rate in the laboratory can be
obtained from Fermi's Golden Rule and utilizing the fact
that the wave-functions of both the initial nucleus and the final one
vanish rapidly outside the nuclear domain and the electron wave-function
in that domain can be approximated as its value at $r=0$
and the neutrino wave-function by a plane wave normalizes to volume
V, so that $H_{if} = \Psi_e(0) g / \sqrt V \int \Psi^*_{^7Li} \Psi^{}_{^7Be} d\tau
= \Psi_e(0) g M_n/\sqrt V $, where $M_n$ represents the nuclear matrix
element and the resultant capture rate is:

\begin{equation}
\lambda_{EC} = 1/\tau_{EC} = (g^2 M_n^2 /\pi c^3 \hbar^4) E_{\nu}^2 |\Psi_e(0)|^2 
\end{equation}

\noindent In the laboratory capture process, any of the various electron shells
contribute to the capture rate; however the K-shell gives the dominant
contribution. At temperatures inside the sun, e.g. $T_6 = 15$,
nuclei such as $^7Be$ are largely ionized. The nuclei however
are immersed in a sea of free electrons resulting from the ionization
process and therefore electron capture from continuum states is possible
(see e.g., \cite{Bet36}, \cite{Bah69b}).
Since all factors in the capture of continuum
electrons in the sun are approximately the same as those in the
case of atomic electron capture, except for the respective electron
densities, the $^7Be$ lifetime in a star, $\tau_s$ is related
to the terrestrial lifetime $\tau_t$ by:

\begin{equation}
{\tau_{fr} \over \tau_t} \sim {2 |\Psi_t(0)|^2 \over |\Psi_{fr}(0)|^2} 
\end{equation}

\noindent where $|\Psi_{fr}(0)|^2$ is the density of the free electrons
$n_e = \rho/ m_H$ at
the nucleus, $\rho$ being the stellar density.
The factor of 2 in the denominator takes care of the two
spin states of calculation of the $\lambda_t$ whereas the corresponding
$\lambda_{fr}$ is calculated by averaging over these two orientations.
Taking account of distortions of the electron wave-functions due to
the thermally averaged  Coulomb interaction with nuclei of charge Z
and contribution due to hydrogen (of mass fraction $X_H$) and heavier nuclei,
one gets the continuum capture rate as:

\begin{equation}
\tau_{fr} = { 2|\Psi_t(0)|^2 \tau_t \over (\rho/M_H)[(1+X_H)/2] 2\pi Z \alpha (m_e c^2 /3 kT)^{1/2}}
\end{equation}

\noindent with $|\Psi_e(0)|^2 \sim (Z/a_0)^3/ \pi$. Bahcall et al \cite{Bah69}
obtained for the $^7Be$ nucleus a lifetime:

$$\tau_{fr} (^7Be) = 4.72 \times 10^8 {T^{1/2}_6 \over \rho(1+X_H)} \rm s$$
The temperature dependence comes from the nuclear Coulomb field corrections
to the electron wave-function which are thermally averaged. For solar condition
the above rate \cite{Bah69b} gives a continuum capture lifetime of
$\tau_{fr}(^7Be) = 140 d$ as compared to the terrestrial mean life of $\tau_t =
76.9 d$.  Actually, under stellar conditions, there is a partial
contribution from some $^7Be$ atoms which are only partially ionized,
leaving electrons in the inner K-shell. So the contributions of
such partially ionized atoms have to be taken into account.
Under solar conditions the K-shell electrons from partially ionized
atoms give another 21\% increase in the total decay rate. Including this,
gives the solar lifetime of a $^7Be$ nucleus as: $\tau_{\odot} (^7Be) = 120 d$.
In addition, the solar fusion reactions have to be corrected for
plasma electrostatic screening enhancement
effects. For a recent discussion of the issues see \cite{Bah02,Qua09}.

\subsubsection{Capture reaction leading to $^8B$}

Apart from the electron capture reaction, the $^7Be$ that is produced
is partly consumed by proton capture via: $^7Be (p, \alpha) ^8B$ reaction.
Under solar conditions, this reaction happens
only $0.02 \%$ of the time. The proton capture on $^7Be$ proceeds at
energies away from the 640 keV resonance via the direct capture process.
Since the product $^7Li$ nucleus emits an intense $\gamma$-ray flux of 478 keV,
this prevents the direct measurement of the direct capture to
ground state $\gamma$-ray yield. The process is studied indirectly
by either the delayed positron or the breakup of the
product $^8B$ nucleus into two alpha particles.
This reaction has a weighted average $S(0) = 0.0238 \rm \; keV barn$
\cite{Fil83}.
The $^7Be (p, \alpha) ^8B$ reaction cross section measurement
has been attempted both by direct capture reactions as well as
by the Coulomb dissociation of $^8B$. For a comparison of the
$S_{17}(0)$ factors determined by the two methods and a critical
review of the differences
of direct and indirect methods, see \cite{Gai06}.

The product $^8B$ is a radioactive nucleus with a lifetime
$\tau = 1.1$ s: 

\begin{equation}
^8B \rightarrow ^8Be + e^+ + \nu_e 
\end{equation}

\begin{figure}[htb]
\centering
\hspace*{-1.3cm}
\includegraphics[height=9cm]{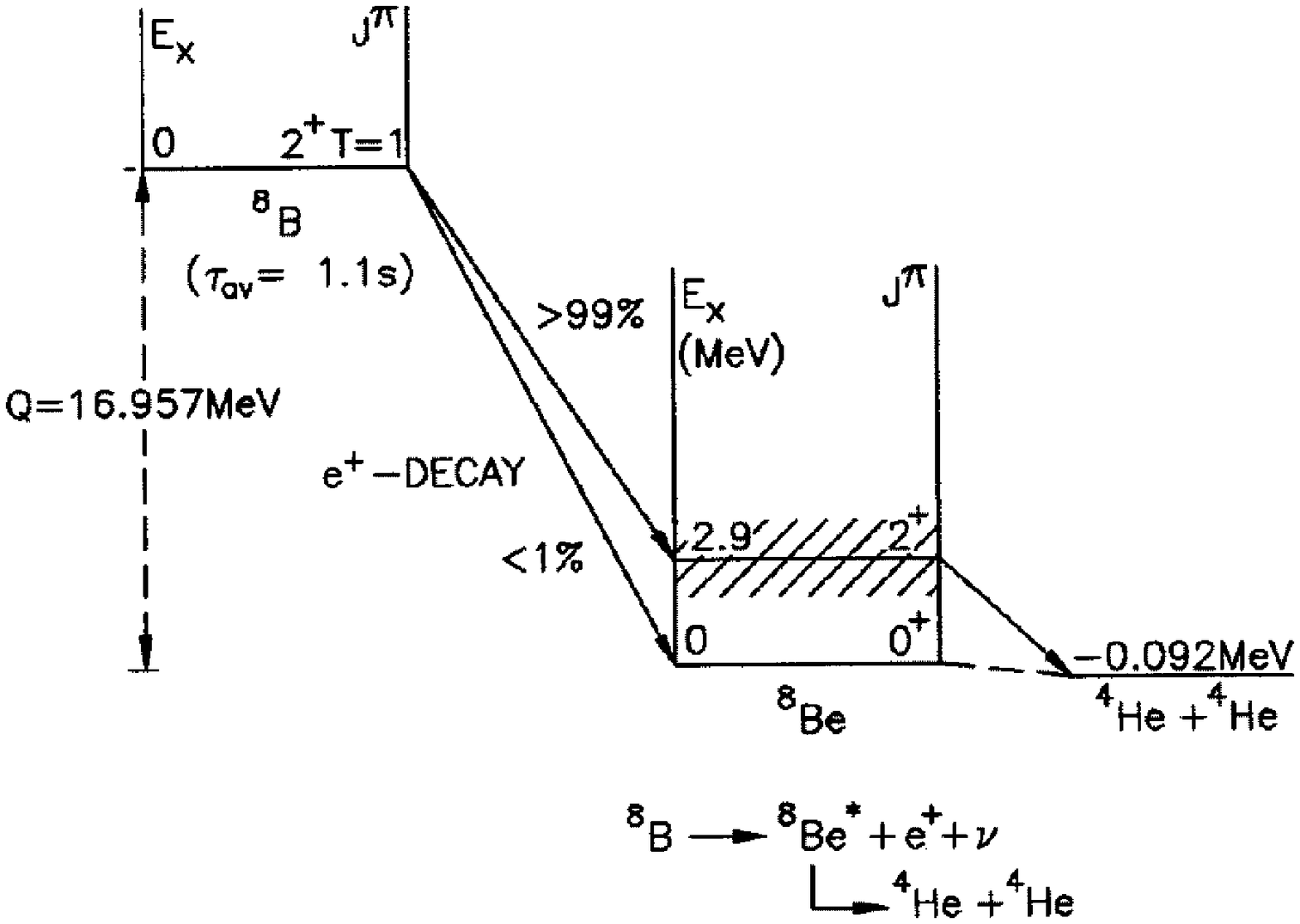}
\caption[]{
The decay scheme of $^8B$ with positron emission, which goes
to the first excited state of $^8$Be at $E_X = 2.9 \; \rm MeV$
with a width of $\Gamma = 1.6 \; \rm MeV$. The $^8Be$ nucleus
itself fissions into two alpha particles. The neutrinos accompanying
the positron decay of $^8B$ are the higher energy solar neutrinos
with $\bar E_{\nu} = 7.3 \; \rm MeV$.
}
\label{fig: B8decay}
\end{figure}

The positron decay of $^8B (J^{\pi} = 2^+)$
goes mainly to the $\Gamma = 1.6$ MeV
broad excited state in $^8Be$ at excitation energy $E_x= 2.94$ MeV
($J^{\pi} = 2^+$) due to the selection rules (see Fig. \ref{fig: B8decay}).
This excited state
has very short lifetime and quickly decays into two $\alpha$-particles.
This completes the ppIII part of the pp-chain. The average energy of
the neutrinos from $^8B$ reactions is: $\bar E_{\nu}(^8B) = 7.3 $ MeV.
These neutrinos, having relatively high energy, play an important
role in several solar neutrino experiments. The neutrino spectrum
is not the same as that obtained
from allowed $\beta$-decay approximation and
is affected by the broad $\alpha$-unstable $^8$Be final state.
Winter et al \cite{Win06}
measured the total energy of the $\alpha$-particles emitted following
the $\beta$-decay of $^8B$ and determined the neutrino
energy spectrum corrected for recoil order effects in the $^8$Be final state
and
constructed from the decay strength function.

\section{The CNO cycle and hot CNO}

\begin{figure}[htb]
\centering
\includegraphics[height=9cm, angle=0.5]{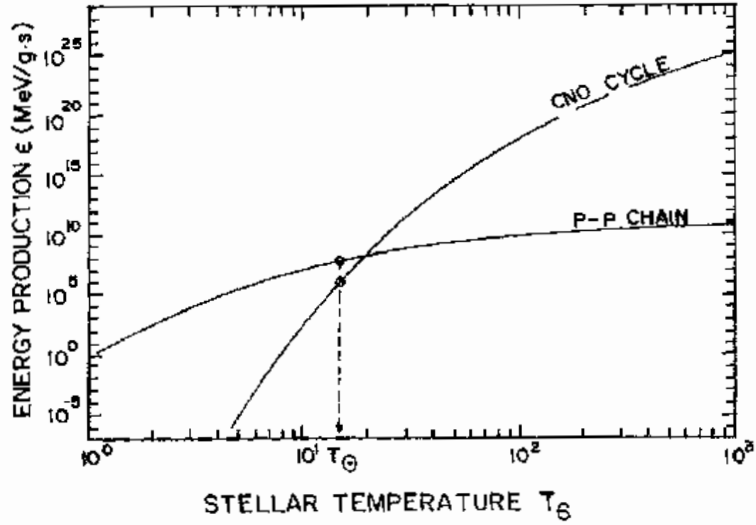}
\caption[]{
Comparison of the temperature dependence of the p-p chain
and the CNO cycle energy production. The points marked
for the solar central temperature $T_{\odot} = T_6 =15.7$
are shown on both graphs. The CNO cycle generation dominates
over the pp-chain at temperatures higher than $T_6 =20$,
so that for sun like stars, the pp-chain dominates. For more
massive stars, the CNO cycle dominates as long as one of
the catalysts: C, N, or O have initial mass concentration
at least 1\%. Note the logarithmic scales of the graph and how
both rates drop sharply with decreasing temperature, with that of
CNO cycle even more drastic due to higher Coulomb barriers.
}
\label{fig: CNOvsPP}
\end{figure}

The sun gets most of its energy generation through the pp-chain reactions
(see Fig. \ref{fig: CNOvsPP}). However, as the central temperature
(in stars more massive than the sun) gets higher, the CNO cycle (see
below for reaction sequence) comes to dominate over the pp-chain at
$T_6$ near 20 (this changeover assumes the solar CNO abundance,
the transition temperature depends upon CNO abundance in the star; in
fact, if one is able to isolate the neutrino flux from the Sun's
weak CN cycle, as in say future extended Borexino or SNO+ experiments with
the capability to detect low energy neutrinos, this could
in turn directly constrain the metallicity of the Sun's 
core\footnote{A measurement CN-cycle neutrino flux (with an expected total
flux of about $5 \times 10^8 \; \rm cm^{-2} \; s^{-1}$)
would test an assumption of the Standard Solar Model
that during the early pre-main-sequence Hayashi phase
the Sun became homogeneous due to convective mixing
and that subsequent evolution has
not appreciably altered the distribution of metals \cite{Hax08}.}\cite{Hax08}.  
The early generation of stars (usually referred to as
the Population II (Pop II) stars, although there is an even
earlier generation of Pop III metal poor massive stars\footnote{Though subsequently
when Carbon has been synthesized by triple-$\alpha$ process, these Pop III
stars turn on their CN cycle.})
generated energy primarily through the pp-chain.
The Pop II stars are still
shining in globular clusters, and being of mass lower than that of the
sun, are very old. Most other stars that we see today
are later generation stars formed from the debris of heavier
stars that contained heavy elements apart from (the most abundant) hydrogen.
Thus in the second and third generation stars (which are slightly
heavier than the sun) where higher central temperatures are possible because
of higher gravity, hydrogen burning can take place through faster chain of
reactions involving heavy elements C, N, and O which have some reasonable
abundance (exceeding 1\%)) compared to other heavy elements like Li, Be, B
which are extremely low in abundance. The favored reactions involve
heavier elements (than those of the pp-chain) which have the
smallest Coulomb barriers
but with reasonably high abundance. Even though the Coulomb barriers
of Li, Be, B are smaller than those of C, N, O (when protons are the
lighter reactants (projectiles)), they lose out due to their lower abundance.

In 1937-1938, Bethe and von Weizs\"acker independently suggested the
CN part of the cycle, which goes as:

\begin{equation}
^{12}C(p,\gamma)^{13}N(e^+\nu_e)^{13}C(p,\gamma)^{14}N(p,\gamma)^{15}O(e^+\nu)^{15}N(p,\alpha)^{12}C
\end{equation}

This has the net result, as before: $4 p \rightarrow
^4He + 2 e^+ + 2\nu_e$ with a $Q=26.73$. In these reactions, the
$^{12}C$ and $^{14}N$ act merely as catalysts as their nuclei are
``returned" at the end of the cycle. Therefore the $^{12}C$ nuclei
act as seeds that can be used over and over again, even though
the abundance of the seed material is minuscule compared
to the hydrogen. But note that there is a loss
of the catalytic material from the CN cycle that takes place through the
$^{15}N (p, \gamma)^{16}O$ reactions. However, the catalytic material is
subsequently returned to the CN cycle by the reaction:
$^{16}O(p, \gamma) ^{17}F(e^+\nu_e)^{17}O(p,\alpha)^{14}N$.

\begin{figure}[htb]
\centering
\hspace*{-0.7cm}
\includegraphics[height=8.5cm,width=11.5cm]{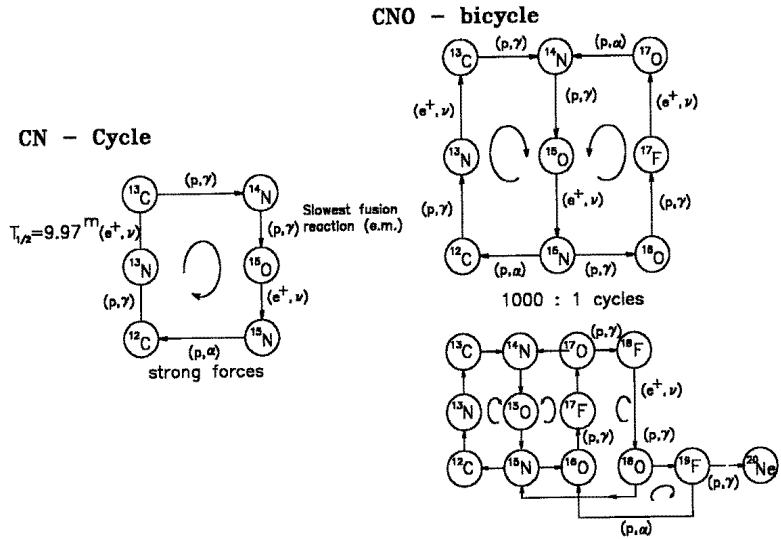}
\caption[]{
The various CNO cycles. The left part is the CN cycle where only
C and N serve as catalysts for the conversion of four protons into
$^4He$. Here the slowest fusion reaction is (p,$\gamma$) reaction
on $^{14}N$ whereas the slower $\beta$-decay has a half-life of
$9.97 \rm m$. In the CNO bi-cycle (right part),
there is leakage from the CN cycle
to the ON cycle through the branching at $^{15}N$. The flow is returned
to the CN cycle (which cycles 1000 times for each ON cycle) through
$^{17}O(p, \alpha)^{14}N$. The right bottom part represents additional
cycles linking into the CNO cycle through the $^{17}O(p, \gamma)^{18}F$
reaction \cite{Rol88}.
}
\label{fig: CNOcycle}
\end{figure}

In the CN cycle (see Fig \ref{fig: CNOcycle}), the two neutrinos involved
in the beta decays (of $^{13}N$ ($t_{1/2} = 9.97 \rm min$)
and $^{15}O$ ($t_{1/2} = 122.24 \rm s$))
are of relatively low energy and most of the total energy $Q=26.73$ MeV
from the conversion of four protons into helium is deposited in the
stellar thermonuclear reactor. The rate of the energy production is
governed by the slowest thermonuclear
reaction in the cycle. Here nitrogen isotopes
have the highest Coulomb barriers in charged particle reactions,
because of their $Z=7$. Among them $^{14}N(p, \gamma)^{15}O$ is the
slowest because this reaction having a final state photon is
governed by electromagnetic forces while that involving the
other nitrogen isotope: $^{15}N(p, \alpha)^{12}C$ is governed by
strong forces and is therefore faster.

From the CN cycle, there is actually a branching off from $^{15}N$
by the reaction $^{15}N (p, \gamma)^{16}O$ mentioned above. This
involves isotopes of oxygen, and is called the ON cycle;
finally the nitrogen is returned to the CN cycle through $^{14}N$.
Together the CN and the ON cycles, constitutes the CNO bi-cycle.
The two cycles differ considerably in their relative cycle-rates:
the ON cycle operates only once for every 1000 cycles of the
main CN cycle. This can be gauged from the S(0) factors of the
two sets of reactions branching off from $^{15}N$: for the
$^{15}N(p, \alpha)^{12}C$ reaction $S(0)= 65 \; \rm MeV \; b$,
whereas for $^{15}N(p, \gamma)^{16}O$, it is $64 \rm \; keV \; b$,
i.e. a factor of 1000 smaller.

\subsection{Hot CNO and rp-process}

The above discussion of CNO cycle is relevant for typical temperatures
$T_6 \leq 80$. These are found in quiescently hydrogen burning stars with
solar composition which are only slightly more massive than the sun.
There are situations where the hydrogen burning takes place at
temperatures ($T \sim 10^8 - 10^9 \rm \; K$) which are in excess
of those found in the interiors of the ordinary ``main sequence" stars.
Examples of these are: hydrogen burning at the accreting surface of a
neutron star or in the explosive burning on the surface
of a white dwarf, i.e. novae, or the outer layers of
a supernova shock heated material in the stellar mantle.
These hot CNO cycles operate under such
conditions on a rapid enough time scale (few seconds) so that even
``normally" $\beta$-unstable nuclei like $^{13}N$ will live long enough
to be burned by thermonuclear charged particle reactions, before
they are able to $\beta$-decay \cite{vWormer94}, \cite{Cha92}. 
So, unlike the normal CNO
the amount of hydrogen to helium conversion in hot CNO is limited by the
$\beta$-decay lifetimes of the proton-rich nuclei like: $^{14}O$ and
$^{15}O$ rather than the proton capture rate of $^{14}N$. Wallace and Woosley
\cite{Wal81} showed that for temperatures, $T \geq 5 \times 10^8 K$,
nucleosynthesised material can leak out of the cycles. This leads to
a diversion from lighter to heavier nuclei and is known as the rapid
proton capture or rp-process.
The flow between the hot CNO cycle and the rp capture process in X-ray bursts from
the atmosphere of a neutron star is
regulated by the $^{15}O(\alpha,\gamma)^{19}Ne$ reaction. 
Another
alpha capture reaction $^{18}Ne(\alpha,\gamma)^{21}Na$, continuously processes  
the available $^4He$ nuclei flowing towards heavier elements. For 
a recent discussion on the hot CNO and the
rp process on accreting neutron stars, see \cite{wie07}.

The nucleosynthesis path of rp-process of rapid proton addition
is analogous to the r-process of neutron addition (for neutron capture
processes in the early galaxy see \cite{sne08}).
The hot hydrogen bath converts CNO nuclei into isotopes near the region
of proton unbound nuclei (the proton drip line). For each neutron number,
a maximum mass number A is reached where the proton capture must wait
until $\beta^+$-decay takes place before the buildup of heavier nuclei
(for an increased neutron number) can take place. Unlike the r-process
the rate of the rp-process is increasingly hindered due to the increasing
Coulomb barrier of heavier and higher-Z nuclei to proton projectiles.
Thus the rp-process does not extend all the way to the proton drip line
but runs close to the beta-stability valley and runs through
where the $\beta^+$-decay rate compares favorably with the proton captures.

\section{Helium burning and the triple-$\alpha$ reaction}

After hydrogen burning in the core of the star has exhausted its fuel,
the helium core contracts slowly. Its density and temperature goes up
as gravitational energy released is converted to internal kinetic
energy. The contraction also heats hydrogen at the edge of the helium core,
igniting the hydrogen to burn in a shell. At a still later stage in
the star's evolution, the core has contracted enough to reach central
temperature density conditions: $T_6 = 100 -200$ and
$\rho_c = 10^2 - 10^5 \; \rm gm \; cm^{-3}$ when the stellar core settles down
to burn $^4He$ in a stable manner.  The product of helium burning
is $^{12}C$. Since in nature, the
$A=5$ and $A=8$ nuclei are not stable, the question arises as to how
helium burning bridges this gap. A direct interaction of three $\alpha$
particles to produce a $^{12}C$ nucleus would seem at first sight, to be too
improbable (as was mentioned, for example, in Bethe's 1939 paper \cite{Bet39},
which was primarily on the
CN cycle). However, \"Opik \cite{Opi51} and Salpeter \cite{Sal52}, \cite{Sal57}
independently proposed
a two step process where in the first step,
two $\alpha$ particles interact to produce
a $^8Be$ nucleus in its ground state (which is unstable to $\alpha$-breakup),
followed
by the unstable nucleus interacting with another $\alpha$-particle
process to produce a $^{12}C$ nucleus.

Thus the triple alpha reaction begins with the formation of $^8Be$
that has a lifetime of only $1\times 10^{-16}$ s (this is found
from the width $\Gamma = 6.8$ eV of the ground state and is the cause of
the $A=8$ mass gap). This is however long compared to the transit time
$1\times 10^{-19}$ s of two $\alpha$-particles to scatter past each other
non-resonantly with
kinetic energies comparable to the Q-value of the reaction namely,
$Q= -92.1 \rm \; keV$. So it is possible to have an equilibrium build-up
of a small quantity of $^8Be$ in equilibrium with its decay or reaction
products: $\alpha + \alpha \rightarrow ^8Be$. The equilibrium concentration
of the $^8Be$ nucleus can be calculated through the Saha equation

\begin{equation}
 N_{12}={N_1 N_2 \over 2} \big({2\pi \over \mu kT}\big)^{3/2} \hbar^3 {(2J +1) \over (2J_1+1) (2J_2 +1)} \rm exp (-{E_R \over k T})
\end{equation}

at the relevant temperature $T_6 =11$ and $\rho = 10^5 \; \rm gm \; cm^{-3}$
to be:

\begin{equation}
{N(^8Be) \over N(^4He)} = 5.2 \times 10^{-10} 
\end{equation}

\begin{figure}[htb]
\centering
\hspace*{-0.5cm}
\includegraphics[height=9.2cm,width=12.1cm, angle=-0.5]{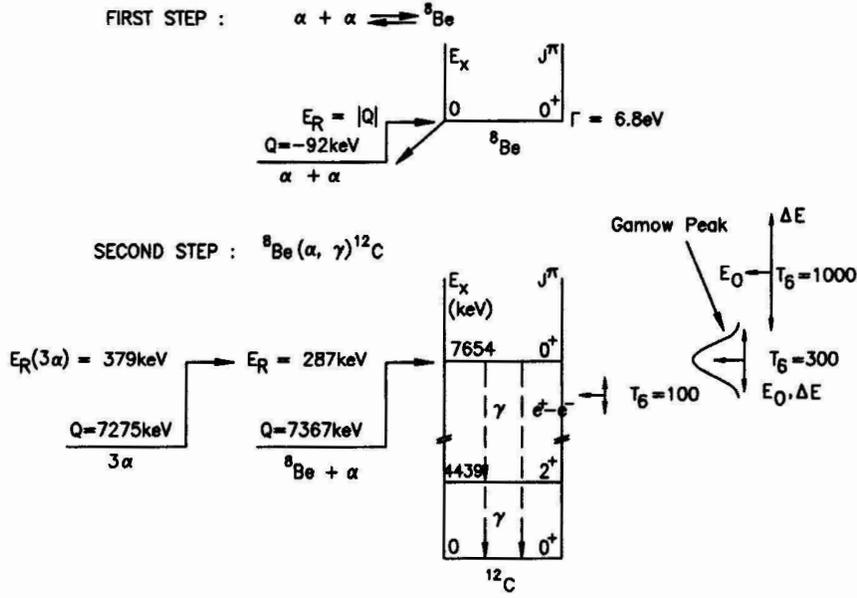}
\caption[]{
Triple alpha process of $^{12}C$ synthesis.
In the first step a small amount of $^8Be$ nuclei builds up
in equilibrium with its decay products (forward and backward
reactions involve alpha particles). The second step involves a
capture of another alpha particle by the unstable $^8Be$ nucleus
which proceeds via an s-wave resonance state in the product
nucleus $^{12}C$ located close to the Gamow energy
window for temperatures indicated schematically
by the three-way arrows on
the right.
}
\label{fig: 3alpha}
\end{figure}

Salpeter suggested that this small quantity of $^8Be$ serves as the seed
for the second stage of the triple $\alpha$-capture into the $^{12}C$ nucleus.
It was however shown by Hoyle \cite{Hoy53}
that the amount of $^{12}C$ produced
for the conditions inside a star at the tip of the red-giant
branch is insufficient to explain the observed abundance, {\it unless} the
reaction proceeds through a resonance process \cite{Hoy54}. The presence
of such a resonance greatly speeds up the rate of the triple-$\alpha$
process which then proceeds through an s-wave ($l=0$) resonance
in $^{12}C$ near the threshold of $^8Be + \alpha$ reaction.
Since $^8Be$ and $^4He$ both have $J^{\pi} = 0^+$, an s-wave resonance
would imply that the resonant state in question has to be $0^+$ in
the $^{12}C$ nucleus.
Hoyle suggested the excitation energy to be: $E_X \sim 7.68$ MeV
in the $^{12}C$ nucleus and this state was experimentally found by
W.A. Fowler's group (\cite{Coo57}) with spin-parity:
$J^{\pi} = 0^+$. This state has a total width (\cite{Rol88})
$\Gamma = 8.9 \pm 1.08$ eV, most of which lies in $\Gamma_{\alpha}$,
due to the major propensity of the $^{12}C$ nucleus to break-up through
$\alpha$-decay. (The decay of the excited state of $^{12}C$  by
$\gamma$-rays cannot go directly to the ground state, since the
resonance state as well as the ground state of the $^{12}C$ nucleus
have both $J^{\pi} = 0^+$ and $0^+ \rightarrow 0^+$ decays are
forbidden. This partial width due to gamma-decay is several thousand
times smaller than that due to $\alpha$-decay).
So, $\Gamma = \Gamma_{\alpha} + \Gamma_{rad} \sim \Gamma_{\alpha}$ and
$\Gamma_{rad} = \Gamma_{\gamma} + \Gamma_{e^+e^-} = 3.67 \pm 0.50$ meV.
Again the radiative width $\Gamma_{rad}$
is dominated by the width due to photon width
deexcitation: $\Gamma_{\gamma} = 3.58 \pm 0.46$ meV. (Note
the scales of {\it milli}electron Volts).
The reaction scheme for the first and the second parts of
the triple-alpha reaction is given in Fig. \ref{fig: 3alpha}.
The locations of the Gamow energy regions near the above resonance state
(for several stellar temperatures) are shown only schematically.

The reaction rate for the $^{12}C$ formation can be calculated
by using the properties of the resonant state and the thermally
averaged cross-section:

\begin{equation}
r_{3\alpha} = N_{^8Be} N_{\alpha} < \sigma v >_{^8Be + \alpha}
\end{equation}

Here $N_{^8Be}$ and $N_{\alpha}$ are the number densities
of interacting $^8Be$ and $^4He$ nuclei and the angular brackets
denote thermal averaging over a Maxwell Boltzmann distribution $\psi(E)$.
This averaging leads to:

\begin{equation}
r_{3\alpha} = N_{^8Be} N_{\alpha} \int^{\infty}_0 \psi(E) v(E) \sigma(E) dE
\end{equation}

with
$$\psi(E) = {2\over \sqrt \pi} {E\over kT} \rm exp (-E/kT) {dE \over (kTE)^{1/2}}$$
and
$$\sigma(E) =  \pi \big({\lambda \over 2 \pi}\big)^2 {2J+1 \over (2J_1+1) (2J_2+1)} {\Gamma_1 \Gamma_2 \over (E-E_R)^2 + (\Gamma/2)^2}$$
is the Breit-Wigner resonant reaction cross section with the resonant
energy centroid at $E=E_R$. The total width
$\Gamma$ is a sum of all decay channel widths such as
$\Gamma_1 = \Gamma_{\alpha}$ and $\Gamma_2 = \Gamma_{\gamma}$.
If the width $\Gamma$ is only a few eVs then the functions $\psi(E)$ and
$v(E)$ can be pulled out of the integral. Then, the reaction rate
will contain an integral like: $\int_0^{\infty} \sigma_{BW} (E) dE
= 2 \pi (\lambda/2 \pi \hbar)^2 \omega \Gamma_1 \Gamma_2/\Gamma$,
where
$\omega = (2J+1) / [(2J_1+1) (2J_2+1)]$
and the functions pulled out of the integral need to be evaluated at $E= E_R$.
Since  most of the time the excited state of the $^{12}C^*$ breaks-up
into $\alpha$-particles, we have $\Gamma_1 =\Gamma_{\alpha}$ dominating
over $\Gamma_{\gamma}$ and $(\Gamma_1 \Gamma_2 / \Gamma) \sim \Gamma_2$.
This limit usually holds for resonances of energy sufficiently high
so that the incident particle width ($\Gamma_1$) dominates the natural
width of the state ($\Gamma_2$). In that case, we can use the number
density of the $^8Be$ nuclei in equilibrium with the $\alpha$-particle
nuclei bath as described by Saha equilibrium condition:

\begin{equation}
N(^8Be) = N_{\alpha}^2 \omega f {h^3 \over (2\pi \mu kT)^{3/2}} \rm exp (-E_r/kT)
\end{equation}

where
f is the screening factor.
It is possible to get the overall triple-alpha reaction rate by
calculating the equilibrium concentration of the excited (resonant) state of
$^{12}C$  reached by the $^8Be + \alpha \rightarrow ^{12}C^*$ reaction
and then multiplying that concentration by the gamma-decay rate
$\Gamma_{\gamma}/\hbar$ which leads to the final product of $^{12}C$.
So, the reaction rate for the final step of the triple-alpha
reaction turns out to be:

\begin{equation}
r_{3\alpha} = N_{^8Be} N_{\alpha} \hbar^2 \bigg({2\pi \over \mu kT}\bigg)^{3/2} \omega f \Gamma_2 \rm exp (-E_r^{'}/ kT)
\end{equation}

where $\mu$ is the reduced mass of the reactants $^8Be$ and $\alpha$ particle.
This further reduces by the above argument to:

\begin{equation}
r_{3\alpha \rightarrow ^{12}C} = {N_{\alpha}^3 \over 2} 3^{3/2} \bigg({2\pi \hbar^2 \over M_{\alpha} kT}\bigg)^3 f {\Gamma_{\alpha} \Gamma_{\gamma} \over \Gamma \hbar} \rm exp (-Q/kT)
\end{equation}

The Q-value of the reaction is the sum of
$E_R(^8Be + \alpha) = 287 \rm \; keV$ and
$E_R(\alpha +\alpha) = |Q| = 92 \; \rm keV$
and turns out to be: $Q_{3\alpha} = (M_{^{12}C^*} - 3 M_{\alpha})c^2
= 379.38 \pm 0.20 \rm keV$ \cite{Nol76}. Numerically, the energy generation rate
for the triple-alpha reaction is:

\begin{equation}
\epsilon_{3\alpha} = {r_{3\alpha} Q_{3\alpha} \over \rho} = 3.9 \times 10^{11}{\rho^2 X_{\alpha}^3 \over T_8^3} f \; \rm exp( - 42.94 /T_8) \rm \; erg \; gm^{-1} \; s^{-1}
\end{equation}

The triple alpha reaction has a very strong temperature dependence: near
a value of temperature $T_0$, one can show that the energy generation rate
is:

\begin{equation}
\epsilon(T) = \epsilon(T_0) ({T \over T_0})^n
\end{equation}

where, $n = 42.9/T_8 -3$. Thus at a sufficiently high temperature and density,
the helium gas is very highly explosive, so that a small temperature rise
gives rise to greatly accelerated reaction rate and energy liberation.
When helium thermonuclear burning is ignited in the stellar core under
degenerate conditions, an unstable and sometimes an explosive condition
develops.

\section{Survival of $^{12}C$ in red giant stars and $^{12}C(\alpha, \gamma)^{16}O$ reaction}

The product of the triple-alpha reactions $^{12}C$, is burned into $^{16}O$
by $\alpha$-capture reactions:

\begin{equation}
 ^{12}C + \alpha \rightarrow ^{16}O + \gamma 
\end{equation}

If this reaction proceeds too efficiently in helium burning Red
giant stars, then all the carbon will be
burned up to oxygen. Carbon is however the most abundant element in the
universe after hydrogen, helium and oxygen, and the cosmic C/O ratio is
about 0.6. In fact, the O and C burning reactions  and the conversion of
He into C and O take place in similar stellar core temperature and
density conditions. Major ashes of He burning in Red Giant stars are C
and O. Red Giants are the source of the galactic supply of $^{12}C$ and
$^{16}O$.
Fortuitous
circumstances of the energy level structures of these alpha-particle
nuclei are in fact important for the observed abundance of oxygen and
carbon.

\begin{figure}[htb]
\centering
\includegraphics[height=8.5cm,width=11.8cm]{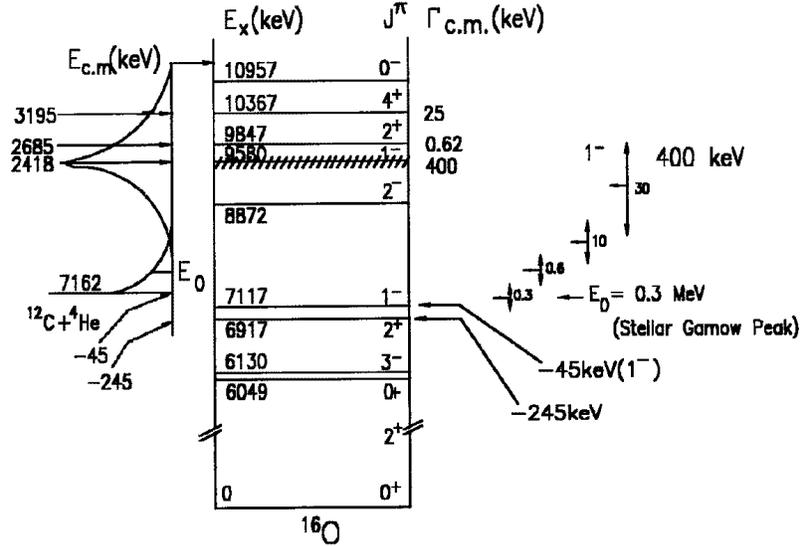}
\caption[]{
Energy levels of $^{16}O$ nucleus near and above the
alpha-particle threshold of capture on $^{12}C$. Shown on
the right are effective stellar energy regions corresponding
to the temperatures given near the three-way arrows. The
reaction rate is influenced mainly by the high energy tails
of two sub-threshold resonances in $^{16}O$ at $E_R = -45 \; \rm keV$
and $E_R = -245 \; \rm keV$, plus the low energy tail
of another high-lying broad resonance at 9580 keV.
}
\label{fig: 16Olevels}
\end{figure}

For example, if as in the case of the 3$\alpha$ reaction, there was a
resonance in the $^{12}C(\alpha, \gamma)^{16}O$ reaction near the
Gamow window for He burning conditions (i.e. $T_9 \sim 0.1 - 0.2$), then
the conversion of $^{12}C \rightarrow \; ^{16}O$ would have proceeded at a
very rapid rate.
However, the energy level diagram of $^{16}O$ shows that for temperatures
upto about $T_9 \sim 2$, there is no level available in $^{16}O$ to foster
a resonant reaction behavior (Fig. \ref{fig: 16Olevels}). But since this
nucleus is found in nature, its production must go through
either: 1) a non-resonant direct capture reaction or 2) non-resonant
captures into the tails of nearby resonances (i.e. sub-threshold reactions).
In Fig. \ref{fig: 16Olevels}, also shown on the left of the $^{16}O$
energy levels, is the threshold for the $^{12}C + ^4He$ reaction, drawn at
the appropriate level with respect to the ground state of the $^{16}O$ nucleus.
The Gamow energy regions drawn on the extreme right for temperatures
$T_9 = 0.1$ and above, indicates that for the expected central temperatures,
the effective stellar (center of mass) energy region is near $E_0 =0.3$ MeV.
This energy region is reached by the low energy tail of a broad resonance
centered at $E_{CM} = 2.42$ MeV above the threshold (the $J^{\pi} = 1^-$ state
at 9.58 MeV above the ground state of $^{16}O$) with a (relatively large)
resonance width of 400 keV.
On the other hand, there are two sub-threshold resonances in $^{16}O$
(at $E_X = 7.12$ MeV and $E_X= 6.92$ MeV), i.e. -45 keV and -245 keV
{\it below} the $\alpha$-particle threshold that have
$J^{\pi} =1^-$ and $J^{\pi} =2^+$, that contribute to
stellar burning rate by their high energy tails.
However, electric dipole (E1) $\gamma$-decay of the
7.12 MeV state is inhibited by isospin selection rules.
Had this not been the case, the $^{12}C(\alpha, \gamma)^{16}O$
reaction would have proceeded fast and $^{12}C$ would have been
consumed during helium burning itself.
The two sub-threshold states at $-45$ keV and $-245$ keV
give contributions to the astrophysical S-factor of:
$S_{1^-}(E_0) = 0.1 \; \rm MeV \; barn$ and $S_{2^+} (E_0) = 0.2 \; \rm MeV \; barn$
respectively at the relevant stellar energy $E_0 = 0.3 \; \rm MeV$.
The state at $E_{CM} = 2.42$ MeV ($J^{\pi} = 1^-$ state at 9.58 MeV)
gives a contribution: $S_{1^-}(E_0) = 1.5 \times 10^{-3} \rm MeV \; barn$.
The total S-factor at $E_0 = 0.3 \; \rm MeV$ is therefore close to
$0.3 \; \rm MeV \; barn$. These then provide low enough S or cross-section
to not burn away the $^{12}C$ entirely to $^{16}O$, so that
$C/O \sim 0.1 $ at the least.

Additionally, $^{16}O$ nuclei are not burnt away by further $\alpha$-capture
in the reaction:

\begin{equation}
^{16}O + ^4He \rightarrow \; ^{20}Ne + \gamma 
\end{equation}

\begin{figure}[htb]
\centering
\hspace*{-0.6cm}
\includegraphics[height=11.2cm,width=12.3cm]{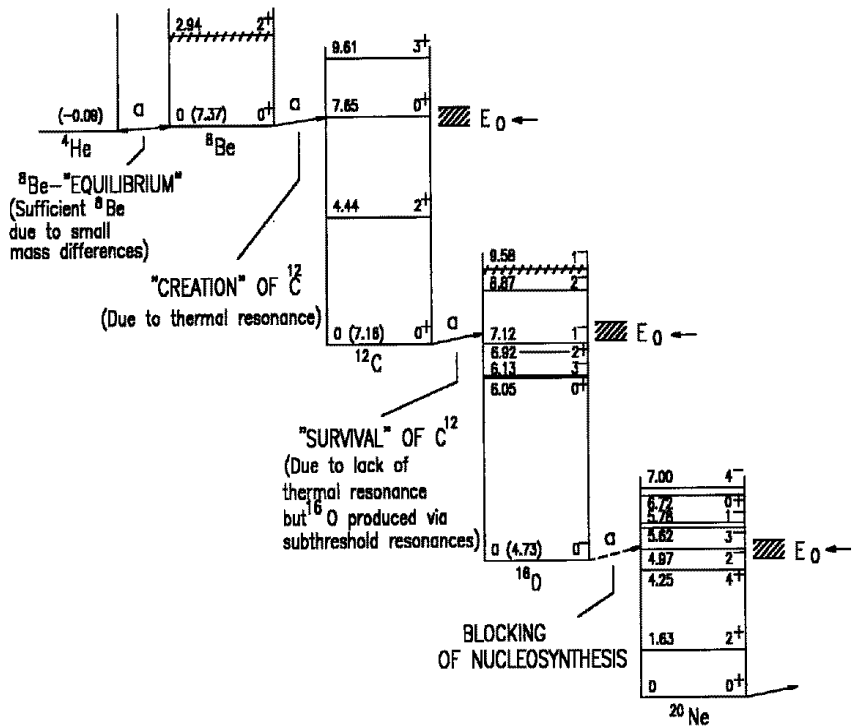}
\caption[]{
Energy levels of nuclei participating in thermonuclear reactions
during the helium burning stage in red giant stars (after \cite{Rol88}).
Survival
of both $^{12}C$ and $^{16}O$ in red giants, from which 
terrestrial abundances result, depends upon the fortuitous
circumstances of nuclear level structures and other properties
of these nuclei.
}
\label{fig: Ne20photo}
\end{figure}

A look at the level schemes of $^{20}Ne$  (see Fig. \ref{fig: Ne20photo})
shows the existence of a
$E_X = 4.97 \; \rm MeV$ state ($J^{\pi} = 2^-$) in the Gamow window.
However, this state cannot form in the resonance reaction
due to considerations of parity conservation (unnatural parity of the
resonant state)\footnote{Whether or not a resonant state can be
formed or accessed via a given reaction
channel depends upon the angular momentum and parity conservation laws.
The spins of the particles in the entrance channel, $j_1, j_2$ and relative
angular momentum $l$ adds upto the angular momentum of the resonant state
$J = j_1 + j_2 +l$. Therefore, for spin-less particles like the closed shell
nuclei $^4He, ^{16}O$ ($j_1 =0, j_2 =0$), we have $J=l$. In the entrance
channel of the  reacting particles, the parity would be:
$(-1)^l \pi(j_1) \pi(j_2) = (-1)^{l=0} (1) (1)$. If the parity of the
resonance state were the same as that of the entrance channel, then
the resultant state would have been a ``natural parity" state. However,
since the 4.97 MeV state in $^{20}Ne$ has an assignment: $J^{\pi} =2^-$,
this is an ``unnatural parity" state.
}. The lower 4.25 MeV state ($J^{\pi} = 4^+$)
in $^{20}Ne$ also cannot act as a sub-threshold resonance as it lies too far
below threshold and is formed in the g-wave state. Therefore only
direct capture reactions seem to be operative, which for $(\alpha, \gamma)$
reactions lead to cross-sections in the range of nanobarns or below.
Thus the destruction of $^{16}O$ via: $^{16}O(\alpha, \gamma)^{20}Ne$
reaction proceeds at a very slow rate during the stage of helium
burning in Red Giant stars, for which the major ashes are carbon and oxygen
and these elements have their galactic origin in the Red Giants.

To summarize, the synthesis of
two important elements for the evolution of life as
we know on the earth have depended upon fortuitous circumstances of
nuclear properties and selection rules for nuclear reactions.
These are: 1) the mass of the unstable lowest (ground) state of $^8Be$
being close to the combined mass of two $\alpha$-particles; 2) there is
a resonance in $^{12}C$ at 7.65 MeV which enhances the alpha addition
reaction (the second step); and 3) parity conservation has protected $^{16}O$
from being destroyed in the $^{16}O (\alpha, \gamma)^{20}Ne$ reactions
by making the 4.97 MeV excited state in $^{20}Ne$ of unnatural parity.

The experimental determination of the reaction rate $^{12}C(\alpha, \gamma)^{16}O$
has been an important goal in nuclear astrophysics for several decades. Its cross
section at the position of the Gamow window for a typical stellar temperature
of $2.5 \times 10^8 \rm K$ is comparable to that of weak interaction cross-sections.
At those energies, this reaction is practically a non-resonant reaction and its
cross-section is determined by the tails of interfering resonance and sub-threshold
states \cite{ang09}. The low cross section and the complexity of low energy
contributions to the reaction rate makes a reliable prediction difficult \cite{buch06,lef08}. 

\section{Advanced stages of thermonuclear burning}

As the helium burning progresses, the stellar core is increasingly
made up of C and O. At the end of helium burning, all hydrogen and helium
is converted into a mixture\footnote{Note however the 
caveat: if the amount of $^{12}C$ is little (either due to
a long stellar lifetime of He burning or due to a larger rate of the
$^{12}C + \alpha \rightarrow ^{16}O + \gamma$ reaction whose estimate
outlined in the earlier section is somewhat uncertain), then the star may
directly go from He-burning stage to the O-burning or Ne-burning stage
skipping C-burning altogether (\cite{Woo86}).
} of C and O, and since H, He
are most abundant elements in the original gas from which the
star formed, the amount of C and O are far more in the core than the traces
of heavy elements in the gas cloud.
Between these two products, the Coulomb barrier
for further thermonuclear reaction involving the products is lower
for C nuclei. At first the C+O rich core is surrounded by He burning shells
and a helium rich layer, which in turn may be surrounded by hydrogen
burning shell and the unignited hydrogen rich envelope.
When the helium burning ceases to provide sufficient power, the star begins
to contract again under its own gravity and as implied by the Virial
theorem the temperature of the helium exhausted core rises. The contraction
continues until either the next nuclear fuel begins to burn at rapid
enough rate or until electron degeneracy pressure halts the infall.

\subsection{Carbon burning}

\begin{figure}[htb]
\centering
\includegraphics[height=11cm,angle=0.2]{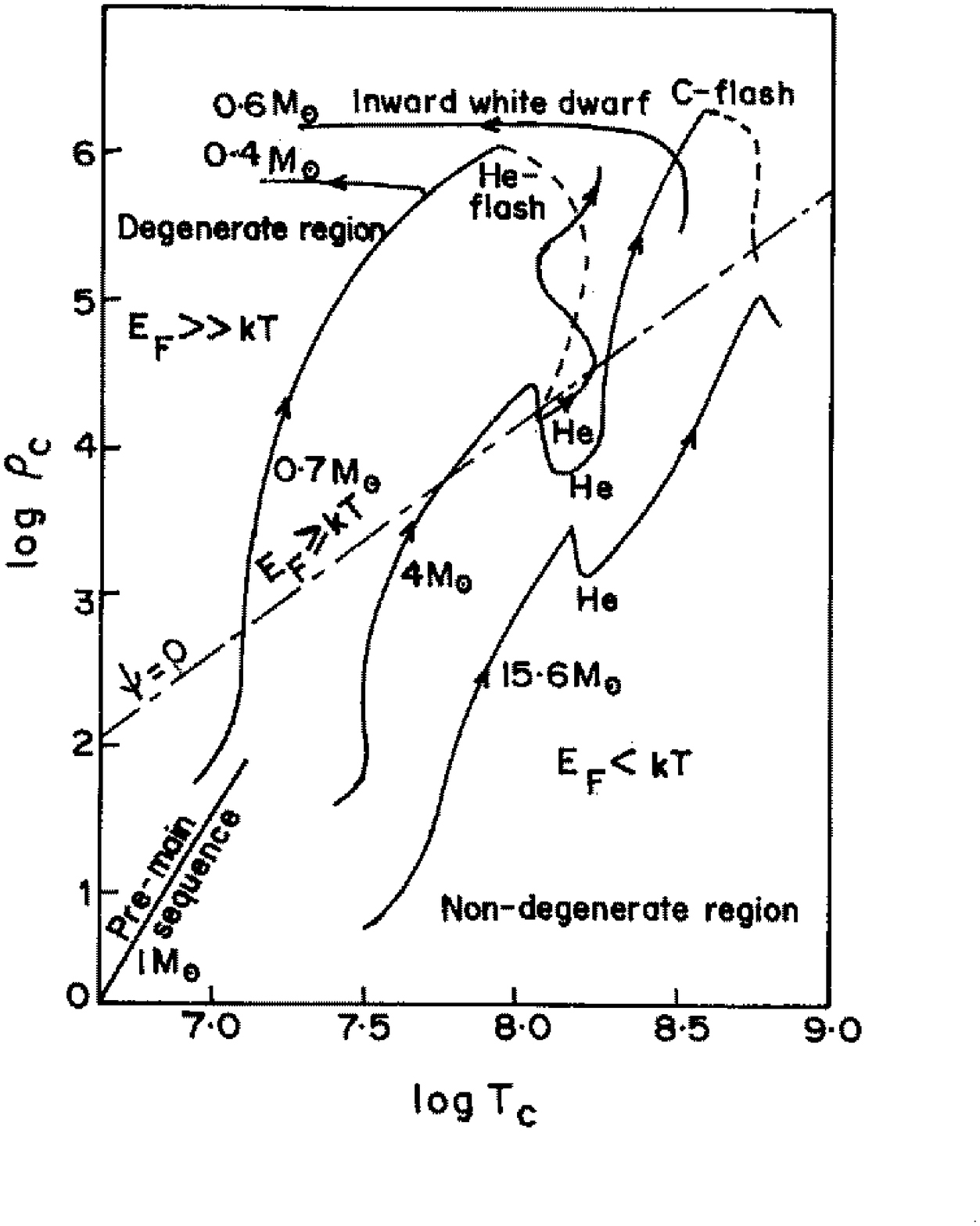}
\caption[]{
Tracks in the core temperature, density plane of
stars of various masses (at the start of hydrogen burning
i.e. main sequence masses). Note that a star of mass
$M \sim 15 M_{\odot}$ ignites all its fuels in
non-degenerate conditions, whereas a star of mass
$M \sim 4 M_{\odot}$ ignites carbon under strongly
degenerate conditions. (After \cite{Hay62}).
}
\label{fig: Hayashi}
\end{figure}

Stars somewhat more massive than about $3 \; \rm M_{\odot}$ contract
until the temperature is large enough for carbon to interact with itself
(stars less massive on the main sequence
may settle as degenerate helium white dwarfs). For
stars which are more massive than $M \geq 8-10 \; M_{\odot}$ (mass on the
main sequence, - {\it not} the mass of the C+O core), the contracting C+O
core remains non-degenerate until C starts burning at
$T \sim 5 \times 10^8 K$ and $\rho = 3 \times 10^6 \; \rm g cm^{-3}$.
Thereafter sufficient power is
generated and the contraction stops and quiescent (hydrostatic, not
explosive) C-burning proceeds (see Fig. \ref{fig: Hayashi}).

The combined mass of two reacting $^{12}C$ nuclei falls at an excitation
energy of 14 MeV in the compound nucleus of $^{24}Mg$. At this energy
there are many compound nuclear states, and the most effective range
of stellar energies (the Gamow window) at the relevant temperature is
about 1 MeV; hence a number of resonant states can contribute to the
decay of the compound nucleus, and even the large angular momentum resonances
may be important because the penetration factors in massive nuclei are not
affected by centrifugal barriers.
The carbon on carbon burning can proceed through multiple,
energetically allowed reaction channels, listed
below:

\begin{equation}
 ^{12}C + ^{12}C  \rightarrow \;^{20}Ne + ^4He \; \; (Q = 4.62 \; \rm MeV) 
\end{equation}

\begin{equation}
 \newline \; \; \; \; \; \; \; \; \; \;    \rightarrow \; ^{23}Na + p \; \; (Q = 2.24 \; \rm MeV) 
\end{equation}

\begin{equation}
\newline \; \; \; \; \; \; \; \; \; \; \;    \rightarrow \; ^{23}Mg + n \; \; (Q = -2.62 \; \rm MeV) 
\end{equation}

At the temperatures where carbon burning starts, the neutron liberating
reactions requires too much particle kinetic energy to be effective.
In addition, based on laboratory measurements at higher energies
compared to the stellar energies,
the electromagnetic decay channel ($^{24}Mg + \gamma$) and the
three particle channel ($^{16}O + 2\alpha$) have lower probability compared
to the two particle channels: $^{23}Na + p$ and $^{20}Ne + \alpha$.
The latter two channels have nearly equal probabilities (see
\cite{Cla68};
at the lowest center of mass energies for which cross-sections
are measured in the laboratory for the proton and $\alpha$ channels,
(i.e. about 2.45 MeV \cite{Maz72}), the branching ratios were
$b_p \sim 0.6$ and $b_{\alpha} \sim 0.4$),
and therefore the direct products of carbon burning
are likely to be $^{23}Na$, $^{20}Ne$, protons and alpha particles.
The rate for this reaction per pair of $^{12}C$ nuclei is (\cite{Ree59}):

\begin{equation}
\rm log \lambda_{12,12} = log f_{12,12} + 4.3 - {36.55(1 + 0.1T_9)^{1/3}\over T_9^{1/3} } -{2\over3} log T_9 
\end{equation}

the factor $f_{12,12}$ is a screening factor. Now, at the temperatures of
$^{12}C$ burning, the liberated protons and alpha particles can be quickly
consumed through the reaction chain: $^{12}C(p, \gamma)^{13}N(e^+\nu_e)^{13}C
(\alpha, n)^{16}O$. Thus, the net effect is that
the free proton is converted into a free neutron (which may be further
captured) and the $\alpha$-particle is consumed with $^{12}C$ into $^{16}O$.
The $\alpha$-particles are also captured by other alpha-particle nuclei,
resulting in, at the end of carbon burning in nuclei like:
$^{16}O$, $^{20}Ne$, $^{24}Mg$ and $^{28}Si$. These secondary reactions augment
the energy released by the initial carbon reaction and Reeves (1959) estimated
that each pair of $^{12}C$ nuclei release about 13 MeV of energy.
Towards the end of carbon burning phase there are also other reactions
such as: $^{12}C + ^{16}O$ and $^{12}C + ^{20}Ne$ which take place.
But these are less rapid and are not expected to play major roles compared
to the $^{12}C+^{12}C$ reactions, due to their increased Coulomb barriers.
A recent discussion of the heavy ion reactions involving C and
O is contained in \cite{Arn96} section 3.6.

During the carbon-burning and subsequent stages, the dominant energy
loss from the star is due to neutrinos streaming out directly
from the stellar thermonuclear furnace, rather than by photons from the
surface. The neutrino luminosity is a sensitive function of core
temperature and quickly outshines the surface photon luminosity of
the star at carbon burning stage. The (thermal) evolutionary time scale
of the star, due to the neutrino emission becomes very short and the core
evolves rapidly, -- so rapidly (compared to the ``cooling" time scale
Kelvin-Helmholtz time: $\tau_{KH} \sim GM^2/R L_{ph}$) that the conditions
in the core are ``uncommunicated" to the surface, since this
happens by photon diffusion. The surface conditions (e.g. the temperature)
then does not change much as the core goes beyond
the carbon burning stage, and it may not be possible just by looking
at a star's surface conditions whether the core is close to a
supernova stage or has many thousands of years of hydrostatic thermonuclear
burning to go.

\subsection{Neon burning}

The result of carbon burning is mainly neon, sodium and magnesium,
but aluminum and silicon are also produced in small quantities
by the capture of $\alpha$, p and n released during carbon burning.
When carbon fuel is exhausted, again the core contracts and its temperature
$T_c$ goes up. At approximately $T_9 \sim 1$, energetic photons from
the high energy tail of the Planck distribution function can begin to
disintegrate the $^{20}Ne$ ash (see Fig. \ref{fig: Ne20photo})
so that one has the reaction: $^{20}Ne + \gamma \rightarrow ^{16}O +^4He$.

Nucleons in a nucleus are bound with typical binding energy of several to $~8$
MeV. An energetic $\gamma$-ray photon is required to photo-eject a
nucleon. Two nucleon ejection requires more energy. Alpha particles
are however released at approximately the same energy as a nucleon due to
the low separation energy of an alpha particle in the nucleus. For example,
the alpha separation energy in $^{20}Ne$ is 4.73 MeV. Thus, the major
photo-nuclear reactions are: $(\gamma, n), (\gamma, p)$ and $(\gamma, \alpha)$
processes. For a photo-disintegration reaction to proceed through an excited
state $E_X$ in the mother, the decay rate is:-
\begin{equation}
\lambda(\gamma, \alpha) = \bigg[\rm exp \big(-{E_X \over kT}\big) {2J_R +1 \over 2J_0 + 1} {\Gamma_{\gamma} \over \Gamma} \bigg] \times {\Gamma_{\alpha} \over \hbar} 
\end{equation}

In the above equation, the first factor in square brackets on the RHS is
the probability of finding the nucleus in the excited state $E_X$ and
spin $J_R$ (with $J_0$ being the ground state spin), while the second factor
$\Gamma_{\alpha} / \hbar$ is the decay rate of the excited state with an
alpha particle emission. Now since $E_X = E_R + Q$, we have:

\begin{equation}
\lambda(\gamma, \alpha) = {\rm exp(- Q/kT) \over \hbar (2J_0 +1)} (2J_R +1) {\Gamma_{\alpha} \Gamma_{\gamma} \over \Gamma} \rm exp(-E_R / kT)
\end{equation}

At $T_9 \geq 1$, the photo-disintegration is dominated by the 5.63 MeV
level in $^{20}Ne$ (see Fig. \ref{fig: Ne20photo}). At approximately
$T_9 \sim 1.5$, the photo-dissociation rate becomes greater than the rate
for alpha capture on $^{16}O$ to produce $^{20}Ne$ (i.e. the reverse
reaction), thus leading effectively to the net dissociation of $^{20}Ne$.
The released $^4He$ reacts with the unspent $^{20}Ne$ and leads to:
$^{4}He + \; ^{20}Ne \rightarrow \; ^{24}Mg + \gamma$. Thus the net result of
the photo-dissociation of two $^{20}Ne$ nuclei is:
$2 \times ^{20}Ne \rightarrow \; ^{16}O + ^{24}Mg$
with a net Q-value of 4.58 MeV. The brief neon
burning phase concludes at $T_9$ close to $\sim 1.5$.

\subsection{Oxygen burning}

At the end of the neon burning the core is left with a mixture of
alpha particle nuclei: $^{16}O$ and $^{24}Mg$. After this another core
contraction phase ensues and the core heats up, until at $T_9 \sim 2$,
$^{16}O$ begins to react with itself:
\begin{equation}
 ^{16}O + ^{16}O  \rightarrow ^{28}Si + ^4He 
\end{equation}

\begin{equation}
\newline \; \; \; \; \; \; \; \; \; \;    \rightarrow ^{32}S + \gamma
\end{equation}

The first reaction takes place approximately $45 \%$  of the time
with a Q-value of 9.593 MeV. In addition to Si and S, the oxygen
burning phase also produces, Ar, Ca and trace amounts of Cl, K, etc
upto Sc. Then at $T_9 \sim 3$, the produced $^{28}Si$ begins to
burn in what is known as the Si burning phase.

\subsection{Silicon burning}

As we have seen, most of the stages of stellar burning involve
thermonuclear fusion of nuclei to produce higher Z and A nuclei.
The first exception to this is neon burning where the photon field is
sufficiently energetic to photo-dissociate neon, before the temperature
rises sufficiently to allow fusion reactions among oxygen nuclei
to overcome their Coulomb repulsion. Processing in the neon burning
phase takes place with the addition of helium nuclei to the undissociated
neon rather than overcoming the Coulomb barrier of two neon nuclei.
This trend continues in the silicon burning phase. In general, a
photo-disintegration channel becomes important when the temperature
rises to the point that the Q-value, i.e. the energy difference
between the fuel and the products is smaller than approximately
$30 k_BT$ (\cite{Hix96}).

With typical Q-values for reactions among stable nuclei above
silicon being 8-12 MeV, photo-disintegration of the nuclear products of
neon and oxygen burning begins to play an important role once
the temperature exceeds: $T_9 \geq 3$. Then nuclei with smaller binding
energies are destroyed by photo-dissociation
in favor of their more more tightly bound
neighbors, and many nuclear reactions involving $\alpha$-particles,
protons and neutrons interacting with all the nuclei in the mass range
$A= 28-65$ take place.
In contrast to the previous burning stages where only a few
nuclei underwent thermonuclear reactions upon themselves, here the nuclear
reactions are primarily of a rearrangement type, in which a particle is
photo-ejected from one nucleus and captured by another and a given fuel
nucleus is linked to a product nucleus by a multitude of reaction chains
and cycles and it is necessary to keep track of many more
nuclei (and many reaction processes involving these) than for previous
burning stages.
More and more stable nuclei form in a nuclear reaction
network as the rearrangement proceeds. Since there exists a maximum in the
binding energy per nucleon at the $^{56}Fe$ nucleus, the rearrangements
lead to nuclei in the vicinity of this nucleus (i.e. iron-group nuclei).

In the mass range $A= 28-65$, the levels in the compound nuclei that form
during silicon burning are so dense that they overlap.
Moreover, at the high temperatures that are involved ($T_9 = 3-5$),
the net reaction flux may be small compared to the large forward
and backward reactions involving a particular nucleus and a quasi-equilibrium
may ensue between groups of nuclei which are connected between separate
groups by a few, slow, rate-limiting reactions (``bottlenecks"). However,
as the available nuclear fuel(s) are consumed and thermal energy is removed
due to escaping neutrinos, various nuclear reactions may no longer occur
substantially rapidly (``freeze-out"). Thielemann and Arnett \cite{Thi85}
found that for cores of massive stars in hydrostatic cases, the bottlenecks
between quasi-equilibrium (QSE) groups coincided with Z=21 nuclei
whereas for lower mass stars, lower temperatures and $Y_e$ and higher density
this bridge involved neutron rich isotopes of Ca capturing protons.
Hix and Thielemann \cite{Hix96}
discussed and contrasted these results with those
of earlier workers; in general the reaction flow across the boundary of
the QSE  groups are influenced by the neutronization of the material,
i.e. the overall $Y_e$. It is in this context that weak interaction processes
such as electron capture and beta decay of nuclei are important, by
influencing the $Y_e$ and thereby the reaction flow. These ultimately
affect both the stellar core density and entropy structures, and it is
important to track and include the changing $Y_e$ of the core material
not only in the silicon burning phase, but even from earlier oxygen burning
phases. The calculation of stellar weak processes on nuclei has spawned
extensive literature (see \cite{Ful82a}, \cite{Kar94}, \cite{nab08} etc., 
and  for a review \cite{Lan02}).

Iron, nickel and neighboring nuclei which are the products of the hydrostatic Si 
burning in the core are mostly trapped in the collapsing core that ends up as
the compact remnant and little of this may reach the interstellar 
medium. However, the blast wave shock launched after the core bounce impacts 
through the onion-like outer shells that the star's earlier evolution has
left behind. The nearest part of the Si shell that is still unburned is heated to 
sufficient temperatures to form iron peak nuclei. Regions which are not as close
can undergo incomplete Si burning and be left with substantial
amounts of Si, S, Ar, Ca and Ti. 
Three separate outcomes may result depending upon the initial density
and the peak temperature: 1) incomplete Si burning; 2) a ``normal freezeout" and 3) an 
``$\alpha$- rich freezeout". In the first case, with initial temperatures of about
$5\times 10^9 \rm K$ and peak density of $\rho = 10^9 \; \rm g \; cm^{-3}$,  
significant amounts of Si and other intermediate-mass elements remain after 
the charged particle reactions freezeout in the expanding ejecta. In the normal
freezeout, an initial condition of $7\times 10^9 \rm K$ and peak density of 
$\rho = 10^9 \; \rm g \; cm^{-3}$ leads to complete Si burning.  
The ``$\alpha$-rich freezeout" however takes place at lower peak densities
$\rho = 10^7 \; \rm g \; cm^{-3}$ though similar peak temperatures: $7\times 10^9 \rm K$. 
This ends up 
with an abundance of $^4 He$ nuclei which produces a significant flow
through the triple alpha reaction into intermediate mass nuclei, in addition to
the iron group nuclei products. The feedback between the rate of nuclear recombination 
and that of temperature evolution in the expansion critically controls the 
``$\alpha$-richness" of matter and production of important radioactive
nuclei e.g. $^{44}Ti$ and is one of the challenges of computational
simulation of silicon burning. For a recent discussion of the computational
aspects of the nuclear evolution during silicon burning, see 
\cite{hix07}. 

In summary, a few key points concerning the thermonuclear burning of $^{28}Si$
are as follows:-

$\bullet$ Direct thermonuclear fusion of two $^{28}Si$ nuclei does not take place
because their Coulomb barrier is too high. Instead thermonuclear fusion takes
place by successive additions of $\alpha$-particles, neutrons and protons.

$\bullet$ Although this is actually a large network of nuclear reactions
it is called ``silicon burning" because $^{28}Si$ offers the largest resistance
to photo-dissociation because of its highest binding energy among intermediate
mass nuclei.

$\bullet$ The source of the $\alpha$-particles which are captured by $^{28}Si$
and higher nuclei is $^{28}Si$ itself. Silicon, sulphur etc. partially
melt-down into $\alpha$-particles, neutrons and protons by photo-dissociation.
These then participate in reaction networks involving quasi-equilibrium clusters
linked by ``bottleneck" links.

$\bullet$ Although beta decay and electron captures on stellar
core nuclei do not produce
energy in major ways they nevertheless play a crucial role in shifting the
pathways of nuclear and thermodynamic evolution in the core conditions.
These ultimately determine the mass of the core and its entropy
structure which finally collapses in a supernova explosion.

\section{Core collapse SNe: electron capture and neutrinos}

At the end of nuclear burning in the core of a massive star, inert Iron group nuclei
are left in the innermost region, -- inert because nuclear burning cannot extract
any further energy from the most tightly bound nuclei. At the last stages of
nuclear burning, neutrino cooling which is far more efficient than the radiation of
photons from the surface, would leave the core compact and degenerate \cite{Chi61}
(where electron Fermi energy is much larger than the thermal energy $kT$) due to a number of
processes such as $e^+ / e^-$ pair annihilation process, photoneutrino process and plasmon
decay into neutrinos. These neutrinos escape from the star freely at this stage since their
interaction cross section is so small, much smaller than that of the photons.
In the picture of the late stages of a spherical star that $\rm B^2FH$ \cite{Bur57} proposed
one had an onion-skin model: nested shells of progressively heavier elements, with the densest
iron core at the center. Successively high-Z elements are ignited in increasingly
central parts of the core in decreasingly lower specific entropy environments, and
are contained within entropy barriers left behind by previous generations of nuclear burning.
The degenerate core becomes unstable to gravitational collapse
at roughly the same Chandrasekhar mass (modulo relatively
small variations in the electron fraction or entropy profiles in the core) 
practically independent of the total mass of the star, which for example could range from
$\sim 8 \rm M_{\odot}$ to $\sim 60 \rm M_{\odot}$ \cite{Woo86}. This ``core convergence"
establishes the important connection between the Chandrasekhar mass and the masses of
neutron stars that are formed from core-collapse supernovae (though the details of the
boundaries of the main sequence masses that lead to neutron stars, as well as the composition
of the cores, e.g. O-Mg-Ne cores vs iron cores etc., have evolved as further research was
undertaken). 
The supernova explosion itself is triggered by the collapse of the core \cite{Gam41},
followed by the sudden release of gravitational energy.
When the iron core grows in size, its temperature increases, and when $T > 7 \times 10^9 K$,
iron nuclei photo-dissociate in endothermic reactions (consuming energy, even though the free 
energy F = U -TS decreases due to increased entropy) into alphas and neutrons. This leads to the 
effective adiabatic index decrease below 4/3 and the iron core becomes unstable to collapse.
Collapse initiation through photo-dissociation happens in relatively massive cores of more
massive stars whereas in less massive stars the collapse may initiate due to reduction
of electrons (which provide the bulk of the supporting pressure
due to their high Fermi momenta) as they are captured by nuclei in ``inverse beta decay" in 
regions where the electron Fermi energy exceed the capture thresholds.

The core collapses on a dynamical scale until the infall is suddenly halted when the 
central density overshoots that of the nuclear matter, 
at $4-5 \times 10^{14} \rm gm\; cm^{-3}$. In between,
the iron core collapsed onto itself nearly freely at about a quarter of the speed of light. 
Initially it had the size of the Earth, but towards the end it is  
a hot, dense, neutron rich sphere about 30 km in radius (see \cite{Woo05} for a review). A static 
accretion shock wave forms initially at the edge of the quasi-free falling core, which soon starts
propagating outward through the outer core and mantle as more kinetic energy is brought into it
by infalling matter. The shock wave soon stalls (at least for most of the range of main sequence
masses beyond about 10 $M_{\odot}$) which may however be helped after about 500 milliseconds
(``after a pause that refreshes" -- according to Hans Bethe) 
\cite{Bet85} by neutrinos which are 
freely streaming out from the inner core, but manage to deposit enough energy (albeit 
a small fraction of their total energy) 
``reviving" the shock.
During this post-bounce phase the proto-neutron star radiates away most of its
energy ($3 \times 10^{53}\; \rm erg$) in the form of neutrinos and antineutrinos of various
flavors and this accounts for the larger intrinsic energy per unit mass of the central engines
of core collapse SNe. 

\subsection{Electron capture on nuclei and protons: a core thermometer}

During the gravitational collapse, the entropy of the core stays low, which permit the nuclei
of various elements present in the core to (largely) survive thermal disintegration and coexist
with a small fraction of ``dripped" nucleons (\cite{Bet79} hereafter BBAL). Around the density
$5 \times 10^{11} \rm gm \; cm^{-3}$, neutrinos produced through electron capture on these nuclei
and free protons are trapped in the core. 
Much of the information pertaining to the conditions in which the
neutrinos are originally produced, such as the nuclear
and thermodynamic properties of the core of the supernova are altered
because these neutrinos undergo {\it inelastic}
scattering with the overlying stellar matter
in the post neutrino trapping phase.
Neutrinos which are emitted through electron captures
on the nuclei present in the pre-supernova and collapsing core {\it before} it
reaches neutrino trapping density ($\simeq 10^{12} \rm \; gm \;cm^{-3}$)\cite{Arn77},
\cite{Bru}, however, stream freely through the stellar
matter without any further interactions. These pre-trapping neutrinos
carry with them information on both physical conditions within the
core, as well as it's nuclear configuration e.g. the ratio of
the number density of free protons to that of 
heavy nuclei.  The last quantity can depend on the nuclear equation of state 
relevant to collapse. 
Since neutrinos act as probes of
the dynamic, thermodynamic and nuclear properties of the pre-supernova and
collapsing core, their detection and measurement of
energy spectrum can have 
significant implications.
The time evolution 
of the detected spectrum could also reveal the dynamical time scale -- a
clue to the average density and mass of the stellar core 
which may have implications for neutron star 
vs black hole formation\footnote{The collapse of a star could in principle
also lead to a ``naked singularity"
\cite{jos09}}. 
The reduction of lepton fraction during stellar
collapse has implications for shock formation stages
and the overall dynamics - 
even in the delayed explosion stage, since it
determines, through the original
energy of the bounce shock, and the entropy profile
in the outer core, the position of the stalled shock.

The loss of neutrinos at low and intermediate densities
is important in determining the saturation ratio of the lepton to baryon ratio at the time of
core bounce (the leptons determine the pressure in the core whereas the baryons, mainly nucleons,
determine the mass of the homologous collapsing core). Brown et al \cite{Bro82} argued that
the hydrodynamic collapse is nearly homologous, i.e. the density structure of the collapsing core
remains self-similar throughout the collapse until the time of bounce. this greatly simplifies
the study of various processes during the collapse and quantities in the core can be calculated
(largely semi-analytically) through the evolution of a ``mean" ``one-zone" symbolizing the core
properties (see e.g. \cite{Ful82b}, \cite{Ray84}). 

The core of a massive star collapses under its own gravity when
the pressure support from degenerate electrons is reduced through
the capture of electrons 
in the stellar material. The electron capture on neutron rich 
heavy nuclei 
initially
proceeds primarily through the allowed type
($\Delta l$ = 0) Gamow Teller transitions. 
As core density exceeds $\simeq 10^{11} \rm gm/cm^3$, the nuclei 
become become more and more massive and too neutron rich to allow 
e$^-$-capture to take place through allowed Gamow-Teller transitions
from the ground state.
This is because the allowed states for $p$ to $n$ transition within
the nucleus are already filled by the neutrons (neutrons shell blocked) 
and the transition strength for
typical captures like $^{56}$Fe $\to$ $^{56}$Mn used earlier (as in
\cite{BBAL}) is
no longer representative of typical nuclear e$^-$-capture rates. 
It was shown  \cite{KR83},\cite{Zar} that the dominant
unique first forbidden transition strength was actually negligible
compared to the thermally unblocked strength under the typical
core collapse conditions. Therefore, after neutron-shell blocking sets in,
(when (A,Z) $>$ $^{74}$Ge) 
the sum rule for the Gamow Teller transition operator $|M_{GT}|^2$
decreases from a typical value of 2.5 \cite{BBAL}, \cite{Ful82b}
to about 0.1.
The e$^-$-capture rate on a single nucleus 
$X(A,Z)$
in the initial state $i$ to the final state j is: 

\begin{equation}
 \lambda_{ij} = \ln 2 {f_i(T,\mu_e, Q_{ij}) \over ft_{ij}}
\end{equation}

\noindent where $ft_{ij}$  is related to $|M_{GT}|^2$  
by $ft_{ij}= 3.596 - \log |M_{GT}|^2$ for 
allowed Gamow-Teller type transitions (for free protons,
log ft$_{f.p.}$ = 3.035).
The factor $f_i(T,\mu_e, Q_{ij})$ is the phase space factor for
allowed transition, which is a function of the 
ambient temperature $T$, the Fermi-energy of the electron $\mu_e$ and 
the Q-value for the reaction.   
The neutrino energy is
$E_{\nu_e} = E_e - Q_{ij}$. 

The change in entropy during collapse controls 
the fraction of the dripped protons with respect to that of the heavy
nuclei and this influences the overall neutrino spectrum received on earth
as the spectrum of neutrinos generated by electron
capture on protons are different from captures on heavy nuclei.
The received neutrino spectrum depends not only upon
the initial conditions from which the collapse started, 
but also 
on the details of the electron capture properties of the stellar
matter. Properties of nuclei at finite temperatures and density
during this phase of the collapse, where shell and pairing corrections
are relevant were computed in \cite{Sut99}
and utilized
to evolve self-consistently with the electron
capture physics and the consequent changes in nuclear and thermodynamic
variables.

The rate of generation of neutrinos per nucleon within energy band
$E_{\nu}$ to $E_{\nu} + d E_{\nu}$ 
after accounting for the relative abundance of free protons and nuclei,
is:

\begin{equation}
 dY_{\nu}(E_{\nu}) = d\lambda_{fp}(E_{\nu}) X_p  + 
                       d\lambda_{H}(E_{\nu}) (1 - X_n - X_p)/A 
\end{equation}

\noindent here A represents the Atomic weight of the ensemble of  nuclei
present in the core, taken to be represented by a single
"mean" nucleus as in 
\cite{BBAL}. The differential neutrino
production rates for free protons and heavy nuclei are:

\begin{equation}
 d\lambda_{fp, H} = { \log 2  \over (ft)_{fp,H} } {<G> \over (m_ec^2)^5}
{ E_{\nu}^2 (E_{\nu} + Q_{fp,H}) \sqrt{ (E_{\nu} + Q_{fp,H})^2 - (m_e
c^2)^2) } \over (1 + \exp( E_{\nu} + Q_{fp,H} - \mu_e)) } d E_{\nu}  
\end{equation}

\noindent where the Coulomb correction factor $<G>$ has been taken
as $\approx 2$  for heavy nuclei and 1 for free protons. 
The Q-value is given as:
 $Q = (\hat \mu + 1.297 + E_{GT})$ 
assuming that the strength is concentrated in a single
state and
here $\hat \mu$ ($=\mu_n - \mu_p$) is the difference in the neutron and 
proton chemical potentials 
when free nucleons coexist with a 
distribution of neutron rich nuclei in nuclear statistical equilibrium, 
and $E_{GT}$ is the energy of the Gamow-Teller Resonance centroid.
The centroids in fp-shell nuclei, found from experimental data
from (n,p) reactions have been used for characterizing GT transitions in
these nuclei \cite{SR95} and are close to the value
(3 MeV) used here. The GT centroid in the e-capture direction is itself a
function of the ambient temperatures as a quasi-particle random phase approximation
(QRPA) calculation shows \cite{Civ99}.
At the relevant high densities, when neutron rich nuclei with $A>65$, 
are abundant and the electron chemical potential 
is noticeably larger than typical nuclear Q-values, 
electron capture rates are mainly sensitive to the centroid and 
total strength of the GT+ distributions -- these are reasonably well described 
within the random phase approximation \cite{Jan07}. 
We note that 
the temperature dependence of the nuclear symmetry energy can
also affect the neutronization of the stellar core prior to neutrino trapping 
during gravitational collapse \cite{Fan08}, \cite{Dea02} since ambient temperatures
can reach upto several MeV.
Not only the reaction Q-values 
but also the equation of state of bulk dense matter, the free nucleon 
abundances, the degree of dissociation into alpha-particles and the nuclear 
internal excitations are modified by changes in the symmetry energy.

\begin{figure}[htb]
\centering
\includegraphics[height=8cm,angle=0.2]{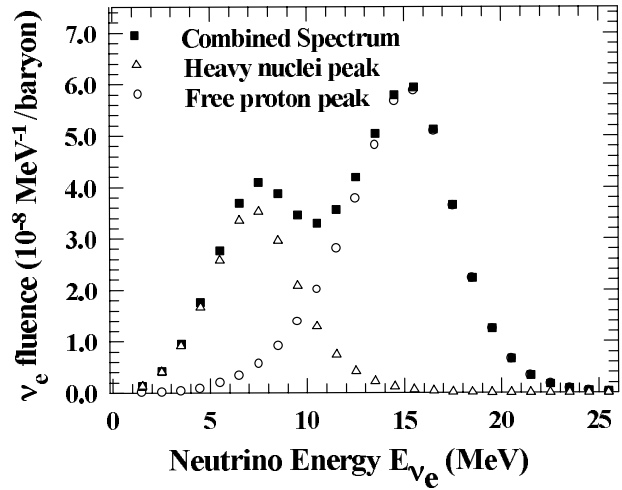}
\caption[]{
``Snapshot" 
neutrino fluence (MeV$^{-1}\; \rm baryon^{-1}$) in a density interval 
$\Delta\rho_{10} = 0.0002$
around  $\rho_{10}=9.8668 \;\rm gm\; cm^{-3}$ for a 15 M$_{\odot}$ star
(D= 1 kpc) 
\cite{Sut97}.
}
\label{fig: neutrino-spectrum-snapshot}
\end{figure}

The Fermi-energy 
$\mu_e = 11.1 (\rho_{10} Y_e)^{1/3} \rm MeV $. The
difference in chemical potentials $\hat \mu$,
and the relative fraction of free protons 
are obtained from a low density analytic equation of state
similar to that in \cite{BBAL} with modifications noted in \cite{Ful82b}.
Shown in Fig. \ref{fig: neutrino-spectrum-snapshot} is a typical ``snapshot"
spectrum of neutrinos from a 15 M$_{\odot}$ star's core collapse
within a narrow
range of stellar core density around $10^{11} \rm\; gm \; cm^{-3}$. 
Note the two separate peaks from
captures on free protons and heavy nuclei
which have non-thermal spectra.

\subsection {Number of neutrinos emitted and predictions of detections}
\label{sec: Number} 

The number of neutrinos and their energy distributions that could have been
detected by large underground neutrino experiments like
the SuperKamioka experiment and the Sudbury Neutrino
Observatory (SNO)\footnote{The original SNO experiment with heavy water has finished taking data.
Thus the heavy water based predictions if a nearby supernova took place are only indicative. 
However there are plans of an extended SNO (called SNO+) which will use 1 kTon of
liquid scintillator which will greatly reduce the lower energy threshold of neutrinos
and will remain sensitive to neutrinos from a nearby SN.
Moreover, after an accident the photomultiplier tubes in SuperKamioka, the low energy thresholds 
originally designed for are no longer operative.} has been calculated \cite{Sut97}.
The spectrally integrated fluence of $\nu_e$
at a distance of 1 kpc, 
as $Y_e$ changes from 0.42 to 0.39 in a 1.4 M$_{\odot}$ stellar core
(of a 15 M$_{\odot}$ star)
is: $F_{\nu} = 4.2 \times 10^{11} \rm cm^{-2}$.
The energy of the infall 
neutrino burst up to this stage is:
$ E_{\nu_e} = 7.2 \times 10^{50} \rm erg $.
The flux, direction and the spectra of the 
neutrinos could have been measured by
the charge current dissociation of the deuterium nucleus 
($\nu_e(d,pp)e^-$) in 
the (``classic") Sudbury Neutrino Observatory (SNO) \cite{SNO} 
with a fiducial mass of 1 Kt
of high purity $\rm{ D_2 O}$.
It would have been also possible to detect $\nu_e$  and obtain spectral information 
by means of the neutrino-electron scattering reaction 
($\nu_e +e^- \to \nu_e+ e^-$) in SNO as well
as in the light water detector Super Kamioka. 
Apart from the reactions mentioned above which can
perform spectroscopy (i.e. measure the incoming neutrino energy),
the neutral current dissociation of $d$ 
by the reaction $\nu_e(d,pn)\nu_e$ could have obtained the 
the total neutrino flux (of all flavors). 
The number of $\nu_e$ events which could have been detected in the SNO  
detector through neutrino-induced 
c.c.  reaction on the target $d$ nuclei  is given by:
 $n_{\nu_e} = F_{\nu} \sigma_{cc}(\epsilon_{\nu_e}) N_{d}$
where $N_d$ $(=6.02 \times 10^{31})$ is the total number of target nuclei 
present in the $1$Kt detector.
The charge current  and neutral current cross-sections 
($\sigma_{cc}(\epsilon_{\nu_e})$ and $\sigma_{nc}(\epsilon_{\nu_x})$ 
respectively),  
have been computed 
for the $\nu$-$d$ process
by Bahcall et al. \cite{Bah88}
and accurate fits to these cross-sections between 5 to 40 MeV
are given as \cite{Burr}:
$ \sigma_i = \alpha_i ( \epsilon_{\nu} - \epsilon_{th, i})^{2.3} $
where $i=cc$ and $nc$, $\alpha_{cc} = 1.7 \times 10^{-44}$ cm$^2$,
$\alpha_{nc} =0.85 \times 10^{-44}$ cm$^2$, $\epsilon_{th,cc} =
2.2$ MeV and $\epsilon_{th,nc} = 1.44$ MeV.

For the H$_2$O based Cerenkov detector
(Super-Kamioka) the ($\nu_e, e^-$) scattering events would be 
the main source of $\nu_e$ spectral information since the corresponding
energy thresholds for charge current and neutral current interactions for
ordinary water  are much higher.
(During the collapse stage, the neutrino flux is almost entirely
in neutrinos of the electron type; 
anti-neutrinos of various kinds, as well
as neutrinos of the mu or tau type are generated in copious numbers
only in the hot post core bounce phase).
The relevant ($\nu_e, e^-$) scattering cross-section is \cite{Seh74}:
$ \sigma_e = (1/2)(4G^2 m_e^2 \hbar^2 / \pi c^2) (7/12)
(\epsilon_{\nu} / m_e c^2) $.

The number of detections by Super-Kamioka (mass 32 kt) and
SNO  for a supernova explosion 1 kpc away, for several possible
scenarios of stellar core collapse are reported
in Table 2
\cite{Sut97}. 
The 15 M$_{\odot}$ star collapse is initiated from thermodynamic
conditions as in \cite{Ful82b} (Y$_{ei}$ = 0.420, $S_i/k_B$ = 1.00,
T$_i$ = 0.7178, $\rho_{10}$ = 0.37), while the 25 M$_{\odot}$ star's
single zone initial conditions are similarly derived from the data 
reported in \cite{WWF} and an expression
for the core averaged entropy (Y$_{ei}$ = 0.423, $S_i/k_B$ = 1.14,
T$_i$ = 0.6756, $\rho_{10}$ = 0.15).
\begin{table*}
\begin{center}
\caption{Pre-trapping
neutrino detections predicted 
in SNO (heavy water) and Super Kamioka \cite{Sut97} with hardness ratios
up to $\rho_{10}$ = 24.16 for indicated heavy nuclear e-capture
matrix elements for 15 M$_{\odot}$ Fuller (1982) and 25 M$_{\odot}$
WWF presupernova stars. Note the caveat in footnote 20.}
\begin{tabular}{c|c|c|cc|cc|cc}
\hline\noalign{\smallskip}
Star Mass  &$|M_{GT}|^2  $ &$t_{collapse}   $ &
\multicolumn{2}{c|} {Pre-trapping Variables}&
\multicolumn{2}{c|} {No. Detected} &
\multicolumn{2}{c|}{Hardness Ratio$^{\dagger}$} \\
& & (ms) & $Y_{ef}$& $S_f/k_B$ &
SNO & S-K & SNO & S-K \\
\hline\noalign{\smallskip} 
15 M$_{\odot}$ & 1.2/0.1 &120 & 0.3909   & 1.0021   & 82     &394       &0.2786
&0.8540  \\
               & 2.5/0.1 &120 & 0.3896   & 1.0085   & 66     &344       &0.2876
& 0.9537  \\ 
25 M$_{\odot}$ & 1.2/0.1 &190 & 0.3828   & 1.1080   &120     &566       &0.2878
&0.8319  \\
               & 2.5/0.1 &190 & 0.3813   & 1.1204   & 99     &499       &0.2916
&0.9190 \\
\hline\noalign{\smallskip}
\end{tabular}
\end{center}
$^{\dagger}$ The hardness ratio denotes the number
of neutrino events in the 5 MeV $\leq E_{\nu_e} \leq 12$ MeV
and 12 MeV $\leq E_{\nu_e} \leq$ 25 MeV bands.
\label{tab: neutrino-detections}
\end {table*}

Neutrino spectroscopy of the final state of a star
would be possible
provided that the event occurs at a relatively close distance. 
Although, a priori, these may be rare events, there have been a
number of historical Supernovae, as well as
detected radio pulsars within a distance of about 2 kpc.
There are a number of star forming regions nearby (such as
the Orion complex - about 440 pc away), 
which are sites of potential supernova progenitor.
A detection by underground neutrino experiments
would constrain features of theoretical
calculations of both collapse and explosion era of type II
Supernovae as well as shed light on the 
characteristics of the stellar core. 
The 19 neutrinos detected from SN1987A were most likely to have been
emitted during the post-bounce phase as their total fluence
during the proto neutron star cooling phase 
(at $\simeq 10^{58}$)
is much larger than that during the collapse phase($\simeq 10^{56}$). 

\section{Detected neutrinos from SN 1987A and future neutrino watch}

Gravitational collapse of the core of the massive star
under its own gravity leads to a supernova explosion. These are
extremely energetic explosions where the observable energy in the kinetic
energy of the exploded debris and electromagnetic radiation add up to several
times $10^{51} \; \rm erg$. The actual energy scale is typically
$3\times 10^{53} \; \rm erg$ or higher, but most of this is radiated away neutrinos. 
Although the full understanding of the process of explosion
in a gravitational collapse leading to a supernova has not been achieved
despite several decades of theoretical and computational work, a watershed
in the field was achieved observationally
when a supernova exploded close by in a satellite galaxy
of our own, namely SN1987A in the Large Magellanic Cloud (LMC). A few neutrinos
were detected from this supernova \cite{bio87}, \cite{hir87}, which were the first detections
of the astrophysical neutrinos from outside of our solar system.
By using the energetics of the neutrinos, their approximate
spectral distribution, the distance to the LMC it was possible to show
that the overall energy of the explosion was indeed
$E_T \sim 2-3 \times 10^{53} \rm erg$.

In addition, the duration of the neutrino burst was of the order of a few
seconds as opposed to a few milliseconds, signifying that the observed
neutrinos had to diffuse out of the dense and opaque stellar matter
as predicted theoretically, instead of directly streaming out of the core.
The spectral characteristics also indicated that the object that is
radiating the neutrinos was indeed very compact, roughly of the same
dimensions as that of a protoneutron star. Thus  SN1987A
provided the observational confirmation of the broad aspects of
the theoretical investigation of stellar collapse and explosion.
For a review of the understanding of the
astrophysics of SN1987A, see \cite{Arn89}.

Physicists are now gearing up to detect not only another supernova
in our own galaxy, but by hoping to build very large neutrino detectors, they
aim to detect supernova neutrinos from the local group of galaxies
(\cite{Cas00}, \cite{Cha00}). 
As neutrinos from the supernova travel directly out from the core,
they arrive a few hours ahead of the light flash from the exploding
star, since light can only be radiated from
the surface of the star, and it takes the supernova shock launched at
the deep core several hours to reach the surface. In the case
of SN1987A, this time delay allowed the estimation of the size of the
star that exploded.
Thus some advance warning of the supernova in the optical band 
can be gotten from a ``neutrino watch".
Physicists
have now connected a worldwide array of neutrino detectors through
the Internet (SN Early Warning System or SNEWS\footnote{See the site:
http://hep.bu.edu/$\sim$snnet/}) which will notify
astronomers to turn their optical, UV and other
telescopes to the right direction
when they find a burst of neutrinos characteristic
of a supernova explosion. 

\section{What X-ray spectroscopy reveals about nucleosynthesis in SNe and SNRs}

X-rays from a supernova explosion arise from the interaction of the supersonic
ejecta with the circumstellar medium (CSM). The CSM typically consists of a slow-moving
wind. When the ejecta collides with the CSM,
it creates two shocks: a high-temperature, low-density, forward-shock
ploughing through the CSM (known as  blast-wave shock)
and a low-temperature, high-density, reverse-shock
moving into the expanding ejecta. Initially the X-rays come
from the forward-shocked shell dominated by continuum radiation, but 
after a few days
X-rays arise also from the reverse-shock, which can have 
substantial line emission, thus providing nucleosynthetic
fingerprints of the ejecta.
The temporal evolution of the X-ray luminosity of a supernova
can yield information on the density distribution
in the outer parts of the exploding star  ($\rho \propto r^{-n}$, here
$n$ can be in the range $7-12$, typically $n \sim 10$ for a Blue
Supergiant (BSG) and $n \sim 12$ for a Red Supergiant (RSG)- see
\cite{che03}). These studies  are therefore
of interest to stellar structure and evolution. 

To date thirty-six supernovae have been detected in the
X-ray bands\footnote{See S. Immler's X-ray supernova page at\\ 
http://lheawww.gsfc.nasa.gov/users/immler/supernovae\_list.html
and \cite{imm03}.}. 
Among these the most extensively studied is SN 1987A because it was so bright.
Another extensively studied object, of the supernova remnant (SNR) kind
is Cassiopeia A. A SNR is just an older supernova after it has picked up enough material from
the surrounding medium, which slows it down. The debris of the explosion,
or the ejecta, radiates when it is reheated after an initial cooling, having
been hit by a inward propagating reverse shock.

\subsection{Supernova Remnant Cassiopeia A}

Cassiopeia A (Cas A for short and also known as 3C461 or G111.7-2.1) is the second youngest
SNR in our Milky Way galaxy. It was believed to be the product of a SN explosion in  $\sim 1672$
\cite{fes06} only about 3.4 kpc away \cite{ree95}. The British astronomer John Flamsteed may have
recorded it as a sixth magnitude star 
in 1680, but it may have faded rapidly after
explosion which could have acted against its widespread reportage. It is a shell type SNR which
was rediscovered by radio astronomers Ryle and Smith \cite{ryl48} 
and is the brightest extra-solar radio
object. Cas A is an oxygen-rich SNR with heavy element distribution
typical of a core-collapse SN. Its closeness, young age
and high brightness across the whole electromagnetic spectrum have underscored its importance 
for studying supernovae\footnote{Here we discuss thermal 
emission from Cas A, which directly connects to nucleosynthetic products. Many shell-type
SNRs, including Cas A, also show non-thermal synchrotron X-rays - see \cite{rey08}.}. 
A recent optical spectrum of the original supernova near 
maximum brightness, obtained from scattered light echo more than three centuries after the direct 
light of the explosion was received on Earth, showed it to be a type IIb 
SN \cite{kra08}\footnote{Krause et al
note that even with the 
overall lack of hydrogen emission in most knots and the nitrogen enrichment in the remnant which is
widely interpreted as signatures of the collapse of a Wolf-Rayet star, i.e. a type Ib SN 
\cite{woo93}, Cas A cannot be classified as arising out of a type Ib SN, since Cas A light echo
spectrum does not match well the spectrum of the prototype SN 2005bf.}, 
somewhat like the well studied prototype SN 1993J\footnote{
The collapse of a red supergiant \cite{ald94} in a binary system \cite{rat92} with
the larger star with a mass on the main sequence of $\sim 10$ to 20 led to SN 1993J and
could explain its light curve
and other characteristics \cite{ray93}, \cite{nom93}. Photometry and spectroscopy
at the position of the fading
SN with Hubble Space Telescope and Keck Telescope, a decade after its explosion
showed the signature of a massive star, -- the companion to the progenitor
that exploded. 
While the binary system initially consisted
of e.g. a $15 M_{\odot}$ star and $14 M_{\odot}$ star
in an orbit of 5.8 yr, at the time of the explosion, the primary had a mass of $5.4 M_{\odot}$
(with a helium exhausted core of $5.1 M_{\odot}$) and a secondary which gained mass in
transfer and which we still see today, ended up with $22 M_{\odot}$ \cite{mau04}.
} \cite{mat00}.
The estimated mass loss rate of $2 \times 10^{-5}\; \rm M_{\odot} \; yr^{-1}$ and a
wind velocity of $10\; \rm km\; s^{-1}$ consistent with the 
hydrodynamical state of the Cas A remnant \cite{cheois03}
is also similar to what has been interpreted from radio and X-ray observations of SN 1993J 
\cite{fra96},\cite{cha04}. 

SNR Cassiopeia A has a bright clumpy emission ring with a diameter of about $3'$.
This is associated with the SNR ejecta, whereas at an outer diameter of about $5'$,
there is fainter emission in a filamentary ring which is due to the forward (``blast-wave") 
shock in the circumstellar medium. 
The SNR has a nearly circular appearance, though the ejecta shows a bipolar structure, 
and jets (mainly in the NE, i.e. upper left of the image and a fainter extension in the
SW direction, called the ``counterjet") that were formed by the explosion (detected both in the
optical \cite{fes96} and in the Si line X-ray emission \cite{hwa00}, \cite{hwa04}). The remnant
has been extensively observed in the optical and more recently with a million second exposure
of the Chandra X-ray Observatory \cite{hwa04} (see also XMM-Newton \cite{wil02})
and includes high resolution grating X-ray
spectroscopy with Chandra \cite{laz06} as also the Spitzer Space Telescope using the 
Infrared Array Camera \cite{enn06} and the IR Spectrograph \cite{smi08}.  

Many of Cas A's optically bright knots have been identified as shocked ejecta which is still
visible due to its young age. The optical radiation emitting regions have been classified into
two groups: the so called fast moving knots (or 'FMK's with speeds $4000 \; \rm km \; s^{-1} < v <
15000 \; \rm km \; s^{-1}$) and the slow moving quasi-stationery floculi (or 'QSF's with speeds
$ v < 300 \; \rm km \; s^{-1}$). The FMKs believed to comprise of 
SN ejecta, have H-deficient emission
(lacks in H$\alpha$) which is dominated by forbidden lines of O and S emission \cite{fes01}
while the QSF emission is rich in N, and is believed to originate from the gas lost from the
star before its explosion (circumstellar envelope) which is now hit by the blast wave shock 
\cite{laz06}. 

Cas A shows the nucleosynthesis products of both hydrostatic and explosive nucleosynthesis. 
C and O are produced in He burning, Ne and Mg first appears through 
C burning, and O and Al are added with Ne burning \cite{Woo95}. Spectral lines of
heavier elements: Si, S, Ar, 
Ca etc are seen. These are
produced in O burning, with alpha reactions on Mg also contributing, -- they are
from zones where explosive O-burning and incomplete explosive
Si-burning occurs. Fe group elements
are produced in the "Si burning" chain of reactions and result from complete and incomplete 
Si-burning. Much of the layered nucleosynthetic 
structure has been preserved during the explosion process in Cas A, with layers of N-, S-, O-rich
ejecta seen beyond the outer shock \cite{fes01}. However, in the optical bands, ``mixed emission 
knots" showing N and S lines suggest that clumps of high speed S ejecta have penetrated through 
outer N-rich layers \cite{fes01}. X-ray line data show that for example,
in the southeast region of the SNR, Fe is farther out and is moving faster than the Si/O region.
Hughes et al \cite{hug00} consider the high surface brightness knots enriched in Si and S are
consistent with nucleosynthetic products of explosive O burning, whereas elsewhere, the more
diffuse, lower surface brightness features with enhanced Fe could be the result of explosive Si
burning.
\begin{figure}[htb]
\centering
\includegraphics[height=7cm]{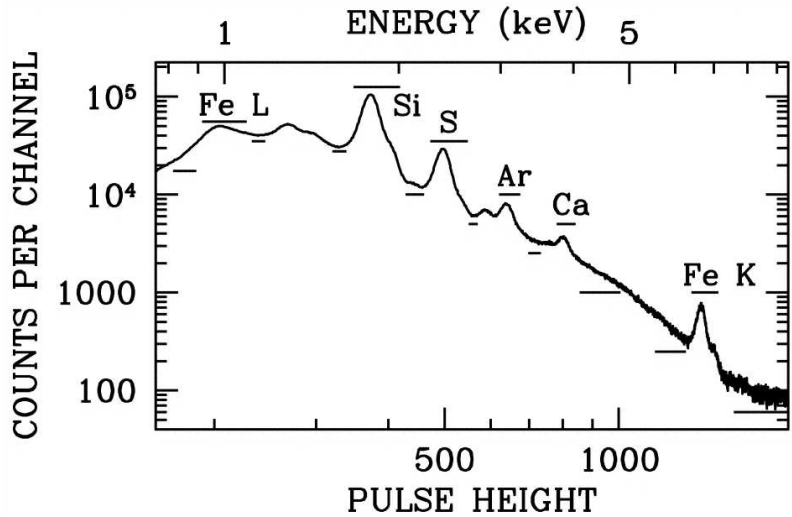}
\caption[]{Spectrum of the entire supernova
remnant Cassiopeia A taken by Chandra (AXAF CCD Imaging Spectrometer (ACIS)) on January 
30-31, 2000 against pulse height (lower x-axis) in the spectral channel and photon energy (upper
x-axis) using a gain value of 4.8 eV per channel (Fig. reproduced by permission of the AAS and 
courtesy of U. Hwang \cite{hwa00}).
}
\label{fig: CasA-ACIS}
\end{figure}

The ``non-dispersed"\footnote{This means the X-ray beam did not pass through any dispersive
element like a 
grating, but the spectrum is obtained by measuring the energy of the incident X-ray 
by the intrinsic spectral sensitivity of the CCD. 
}
X-ray CCD  
spectrum for the entire Cas A remnant is shown in 
Fig. \ref{fig: CasA-ACIS}.  
A 50 ks exposure in January 2000 collected approximately 
16 million photons during the observation of the SNR.
It shows prominent lines of Si, S, Ar, and Ca, in their Helium-like ionic state and undergoing
$(n = 2 \rightarrow n=1)$ transitions as  
also the L and K transitions of Fe.

\begin{figure}[htb]
\centering
\includegraphics[height=9cm]{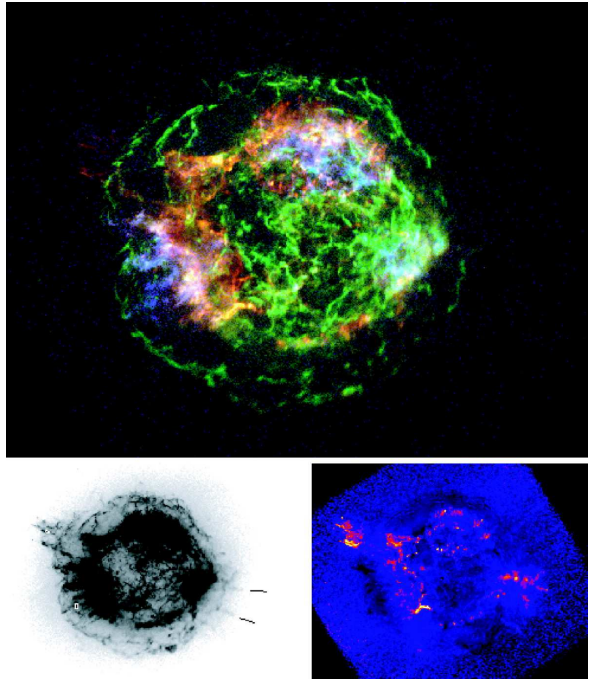}
\caption[]{
Top: Three-color image of Cas A with red = Si He$\alpha$ (1.78–2.0 keV), blue = Fe K 
(6.52–6.95 keV), and green = 4.2–6.4 keV continuum. The remnant is roughly 5' across. 
Bottom left: Overexposed broadband image showing faint features. 
The spectral regions are indicated (top left of this box: northeast jet; 
bottom left of this box: Fe-rich region; 
lines at bottom right point to two southwest jet filaments). 
Bottom right: On the same scale, the ratio image of the Si He$\alpha$ (1.78–2.0 keV) 
and 1.3–1.6 keV (Mg He$\alpha$, Fe L), without subtraction of the continuum contribution. 
The image highlights the jet and counterjet traced by Si emission, although features at the 
lowest intensity levels are uncertain. (Figure reproduced by permission of the AAS and
courtesy of U. Hwang \cite{hwa04}). 
}
\label{fig: CasA-image-million-sec}
\end{figure}
Maps were constructed of the entire remnant in the spectral line energies of several
specific elements.
The Si and S X-ray line maps and the optical maps resemble each other and
this  demonstrated that the X-ray and optically emitting ejecta are largely spatially coexistent. 
In fact the X-ray Si and S line emission, expected to include a significant contribution from the
ejecta, shows asymmetric Doppler shifts corresponding to bulk velocities of $\sim 2000 \; \rm km\;
s^{-1}$ which are comparable to those of the optical ejecta knots \cite{hwa00}.
Si, S, Ar, and Ca have similar general morphologies, but examination of their images in their
line energies also show significant variations. At the positions of the brightest knots in the Si 
image at the inner boundary of the shell to the northeast and southeast the corresponding
Ar and Ca X-ray images are much weaker in the remnant. 
Hughes et al \cite{hug00} using the Chandra first light observation of Cas A ($\sim 5000 s$ on 1999
August 20) showed that the Fe-rich ejecta lie outside the S-rich material, and claimed that
this is due to extensive, energetic bulk motions which caused a spatial inversion of a part
of the supernova core. 
However, on the basis of 
Spitzer Space Telescope imaging and spectral data Rudnick and collaborators
\cite{rud08}, \cite{smi08} claim that there are two roughly spherical shocks, (the blast wave and
the reverse shocks) but ``the wide spatial variations in composition seen in 
infrared, optical and X-ray studies are not due to local differences in ionization age or
temperature, but instead reflect very specific asymmetries in the geometry of the underlying
explosion. These asymmetries are ones in which the nucleosynthetic layers have remained mostly
intact along each radial direction, but the velocity profiles along different radial directions
vary over a range of approximately five" (see also \cite{enn06}). 
In some directions, only the upper C-burning layers have been probed by the reverse shock, 
while in other directions, deeper O- and Si-burning layers have reached the reverse shock.

\begin{figure}[htb]
\centering
\hspace*{-0.8cm}
\includegraphics[height=7.5cm]{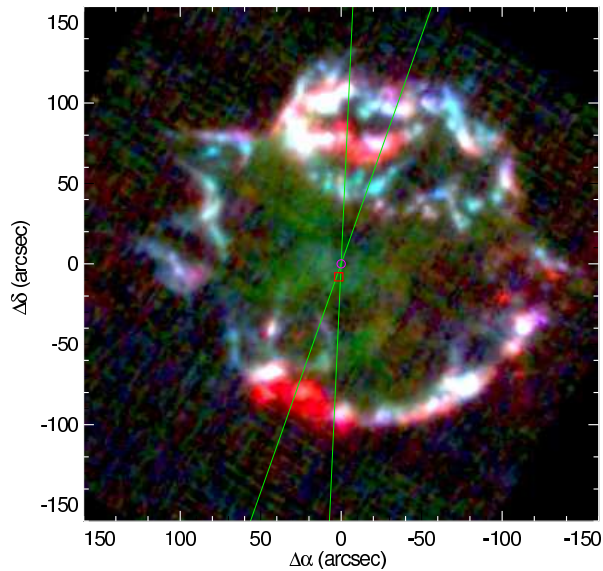}
\caption[]{
Composite Spitzer image of Cas A in [Ar III] (blue), [S IV] (green), and [Ne II]
(red), showing two pronounced neon-rich crescent-shaped regions to the
N and S. 
The size of the image is $320"$ with north to the top and east to the left.
Shown are the kinematic center of the remnant at the magenta circle
and X-ray localization of the remnant's compact object (red square, $7"$ to
the south).  Green lines indicate the 1$\sigma$ range of the ``kick
vector'' direction of the compact object from the ejecta's expansion
center: $169^o \pm 8.4^o$ \cite{fes06}.  The two regions of
enhanced neon abundance lie very close to this projected direction.
Several much smaller neon-enhanced regions lie in the West along the
X-ray jet direction
(Fig reproduced by permission of AAS and courtesy of J. Smith \cite{smi08}). 
}
\label{fig: CasA-Spitzer-08-neon-moons}
\end{figure}
The gas infrared lines of Ar, Ne, O, Si, S and Fe are seen at different locations, along with
higher ionization states of the same elements visible in the optical and X-ray lines. For example,
strong [Ar II] regions in the IR match very closely the places where helium- and hydrogen-like
ionization states of Si and S are seen in the X-ray and [S III] seen in the optical \cite{fes01}.
These are the same regions with multiple temperatures and ionization states but show products of
oxygen burning instead of carbon burning.
Other regions, for example the crescent-shaped green feature slightly east of south in the Spitzer
IRAC line and continuum images depicting gas in Fig 2 of \cite{enn06} show strong [Ne II] emission
in the IR (see also the red regions in the image Fig \ref{fig: CasA-Spitzer-08-neon-moons} 
reproduced from
\cite{smi08}), increasing [O III] emission in the optical and a gap in the silicon dominated X-ray
emission, are interpreted as the locations where carbon-burning layers are
presently encountering the reverse shock. 
Different
layers containing various types of nucleosynthetic products seems to show the presence of 
characteristic types of dust, e.g. the deep layers contain silicates while the upper layers
contain dust dominated by $Al_2O_3$ and carbon grains. They also find evidence 
for circumstellar dust heated by the blast wave shock \cite{rud08}.

Smith et al \cite{smi08} identify IR line emission from ejecta materials in the interior, prior to 
their encounter with the reverse
shock, as well as from the post-shock bright ring. There is a dramatic increase in the electron
density (by a factor $\geq 100$ to $\sim 10^4 \rm cm^{-3}$) as well as a concomitant change in 
the ionization state of the ejecta as it encounters the reverse shock. There is a clear layering 
of ionization state, from low ionization species in the interior, e.g. [Si II] (8.2 eV), higher 
energies on the IR bright ring [S IV] (34.8 eV) where optical
emission is also seen, and very high energies traced by X-ray line emission from H-like and He-like
K-alpha resonance lines of Si XIII (0.5 keV) and Si XIV (2.4 keV), extending beyond the IR-bright
rim. In addition, they find two compact, crescent shaped clumps with high neon 
abundance (mentioned above) which
are arranged symmetrically around the central neutron star, and the crescent regions are closely
aligned with the kick direction of the neutron star from the remnant's expansion center. These
regions contain a huge amount of ionized neon (dominated by Ne$^+$ and Ne$^{++}$, and excluding
neutral neon), almost $1.8 \times 10^{-4} \; \rm M_{\odot}$, flowing outwards $~ 20$ degrees
from the plane of the sky at roughly $-5500 \; \rm km \; s^{-1}$ in the south
and $+4200 \; \rm km \; s^{-1}$ in the north, while the entire SNR may contain
an ionized neon mass $\sim 8.6 \times 10^{-4} \; \rm M_{\odot}$.
Smith et al comment that the {\it apparent} macroscopic elemental mixing mentioned in
\cite{hug00}, may actually arise from different compositional layers of ejecta passing through
the reverse shock at present along different directions.

\subsubsection{X-ray grating spectra of Cassiopeia A and SN 1987A}

In the optical bands, thousands of individual knots have been observed to yield
kinematic information, whereas the X-rays probe more dynamically important information since
a much larger fraction of the ejecta mass is probed by X-rays ($\sim 4 M_{\odot}$ as compared 
to $< 0.1 M_{\odot}$ in the optical). The Chandra High Energy Grating Spectroscopy (HETGS) 
\cite{laz06} provided for the first time the high spatial and spectral resolution 
X-ray map of Cas A, 
comparable to what is achievable in the optical. Because of Cas A's bright emission lines and 
narrow bright filaments and small bright clumps that stand out against the diffuse, continuum 
emission, this grating X-ray spectroscopy was possible and yielded rich information about
the kinematic and plasma properties of the emitting knots. The HETGS has two grating arms 
with different dispersion directions, with the medium energy range (MEG) covering the range
$0.4 - 5.0 \; \rm keV$ and spectral FWHM of 0.023 \AA, while the high energy range (HEG) covers
$0.9 -10.0 \; \rm keV$ and spectral FWHM 0.012 \AA. The grating spectra were obtained 
by Lazendic et al on some
17 different positions on the Cas A image, mainly near the reverse shock regions which are
X-ray bright. 

As mentioned already, Cas A image in the HETGS both in the non-dispersed (zeroth-order)
as well as dispersed images in different energy bands contain H- and He-like ionization
states of O+Ne+Fe (0.65-1.2 keV), Mg (1.25-1.55 keV), Si (1.72-2.25 keV), S (2.28-2.93 keV), 
Ar (2.96-3.20 keV), Ca (3.75-4.0 keV). In addition images in the Li- and Be-like states of 
Fe K lines (6.3-6.85 keV) are also obtained. 
A key advantage of high resolution grating spectroscopy over the low resolution
``non-dispersive" CCD spectroscopy is that the former can be used to resolve individual lines
and thereby interpret conditions in the radiating plasma in various parts of the SNR.
The He-like ions of Si and S and other elements present in Cas A are the dominant ion species
for each element
over a wide range of temperatures and they emit strong K-shell lines of these ionic stages.
For typical plasma densities in the SNR, the He-like triplet of Si and S lines shows
strong forbidden (f) and resonance lines (r) and a comparatively weaker intercombination line (i)
(see section 4.5 of \cite{lie99} for an illustration of these transitions in O-ions). The G-ratio 
$= (f + i)/r$ is, for example, a good diagnostic of the electron temperature \cite{por01}. 
Thus the measured line ratios and abundance ratios of H- and He-like ions of the same element
(say Si) were used \cite{laz06} to measure the electron temperature $kT_e$ and the ionization
time scale $\tau = n_e t$ ($n_e$ being the electron density and $t$ being for example
the time since the region was hit by the shock). The results of \cite{laz06} show that for
most of the selected regions, resolved spectroscopy of Si He-like triplet lines and Si 
H-like lines 
yields  temperatures around $\sim 1 \; \rm keV$, consistent with reverse shocked ejecta. 
However, two regions, (designated R8 and R10) had significantly higher temperatures 
$\sim 4 \; \rm keV$, and could be part of the circumstellar material.  The dominant element in 
Cas A seems to be Oxygen, and the electron density in the X-ray line emitting regions seem to 
vary between 20 to 200 $\rm cm^{-3}$, which is more typical of the unshocked interior of the IR 
line-emitting plasma, than that at the shock front as shown in \cite{smi08} (see above). Lazendic 
et al had derived unambiguous Doppler shifts for their selected 17 regions. While the SE region 
of the SNR show mostly blue shifted velocities reaching upto $-2500 \; \rm km \; s^{-1}$ (compare 
the IR line velocities in the south $-5500 \; \rm km \; s^{-1}$), the NW side of the SNR had 
extreme red shifted values up to $+4000 \; \rm km \; s^{-1}$
(compare IR line velocity in the north: $+4200 \; \rm km \; s^{-1}$).

\begin{figure}[htb]
\centering
\includegraphics[height=5cm]{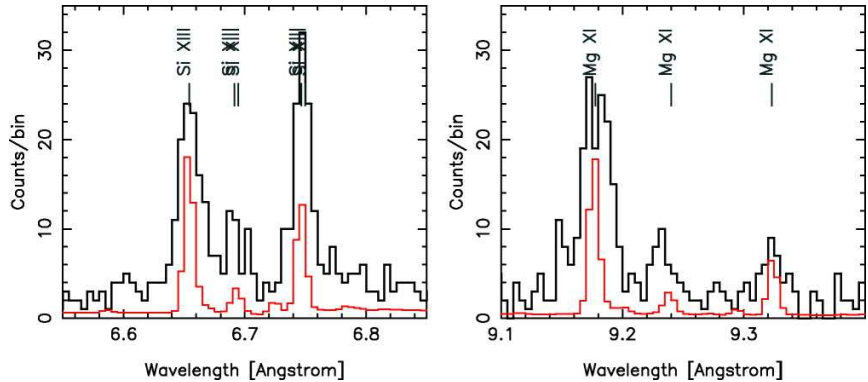}
\caption[]{
Si (left panel) and Mg (right panel) triplets (resonance, intercombination 
and forbidden lines) 
resolved by Chandra High Energy transmission Grating Spectroscopy. 
The data are shown by the solid black histogram and an arbitrarily scaled 
point-source version of the two-shock model (no spatial-velocity effects) 
is shown in red to suggest the relative similarity of the data and model 
G-ratios [(f+i)/r]. (Fig reproduced by permission of the AAS and courtesy 
of D. Dewey \cite{dew08}).
}
\label{fig: HETG-SN1987A-Si-Mg}
\end{figure}

Deep high resolution X-ray spectroscopy of SN 1987A was undertaken with 
Chandra HETG, 20 years after its explosion \cite{dew08}. The impact of the SN ejecta
with the circumstellar medium dominates the observed luminosity, the X-ray luminosity
having brightened by a factor of 25 from that in 1999 and is presently increasing
at the rate of 40 \% per year \cite{par07}. An expanding elliptical ring is seen
in the X-ray image whose brightness distribution seems to correlate with the rapidly
brightening optical hot spots on the inner circumstellar ring observed with Hubble Space
Telescope. The HETG spectrum for a total live time of 355 ksec
shows H-like and He-like lines of Si, Mg, and Ne, and O VIII lines and bright Fe XVII lines.
Fig. \ref{fig: HETG-SN1987A-Si-Mg} taken from Dewey et al \cite{dew08} shows 
the resolved Si and Mg triplets of
(r, i, and f lines); the data have similarity with model G-ratios $(f+i)/r$ mentioned
earlier. Since the dispersed X-ray spectrum is a convolution of the spatial
structure of the X-ray image and the motion of the X-ray emitting gas, Chandra's
sub-arc second resolution was important to resolve the circumstellar ring
of dimensions $1.2" \times 1.6"$.

The global fit to the HETGS data with a two shock model yielded element
abundances and absorption column densities which are consistent within the same
90\% confidence limits that were derived from LETGS by \cite{zhe06}.
The lower temperature shock seems to give the same $kT_{low}$ for all the
LETG data sets (in 2004 and 2007) and the HETG observation, whereas
the $kT_{high}$ values seem to show a general evolution towards lower
values among the grating datasets and similar trend seen in the ACIS monitoring
of SN 1987A. 

The HETG data implies a relatively low bulk radial velocity
of the shocked gas in the ring compared to model expectation from a plane
parallel strong shock entering stationary gas and the temperature
range inferred from the spectral modeling of the emitting gas.
Since the X-ray image is correlated with the optical hot spots, it
appears that the blast wave ahead of the SN ejecta is overtaking dense
clumps of circumstellar gas in the hot spots. The blast wave may be
encountering the dense clumps either at normal incidence or at oblique
incidence. In the former case, the reflected shock would give rise to
gas that has been shocked twice and having nearly stationary bilk velocity
but further elevated temperature. With more and more X-ray emission coming
from gas behind reflected shocks, there would be an increase
of the fraction of higher temperature X-ray emission, consistent with what
is seen. If the blast wave encounters dense clumps at an oblique incidence, the
shocked gas will have significant velocity component parallel to the
shock surface, and a significantly fast transmitted component emitting
X-rays can result.
The Doppler broadening seen in the line profiles of the Chandra data
seems to suggest that both transmitted and reflected shocks encountering
the circumstellar ring at normal and oblique angles may be at play \cite{dew08}. 

\subsection{Live radioactive decays in Cas A, SN 1987A}

Gamma-ray, X-ray, optical and infrared line spectroscopy of SNe and SNRs have been used to observe 
nucleosynthesis and the abundance distribution of elements
freshly synthesized (both radioactive isotopes and stable decay products)
and to extract dynamical information about the explosion. 
In particular the radioactive isotopes provide unique tracers of nucleosynthetic processes
(what, where and how much) and its related dynamics.
The best examples are the observations of gamma-ray lines in supernovae, but X-ray line
spectroscopy
of decay products of radioactive nuclei have also been attempted, and 
specific elements in numerous SNRs and a few SNe identified.

After a few days, the main energy input to the SN ejecta comes from radioactive decay. For
SN 1987A, at
first, the important isotopes are $^{56}Ni$ ($\tau_{1/2} = 6.1 d$) followed by $^{56}Co$
($\tau_{1/2} = 78.8 d$). Beyond 
$\sim 1100$ days, $^{57}Co$ is
more important and at $\geq 2000 d$, the dominant role is played by $^{44}Ti$.
Using bolometric and broad-band UBVRI light curves \cite{frakoz02} estimated the masses
of the three most important radio-isotopes in SN 1987A to be 
$^{56}Ni (0.069 \pm 0.003 M_{\odot})$, 
$^{57}Ni (0.003 M_{\odot})$ and $^{44}Ti (1 \pm (0.5 - 2) \times 10^{-4} M_{\odot})$. 

Among the radioactive isotopes accessible to gamma-ray astronomy, $^{44}Ti$ is a key isotope
for the investigation of the inner regions of core collapse SNe and their young SNRs. It has a
half-life of $58.9 \pm 0.3 \; \rm yr$ \cite{ahm06} and a decay scheme: 
$^{44}Ti \rightarrow ^{44}Sc \rightarrow ^{44}Ca$. The discovery \cite{iyu94}
of the 1157 keV gamma-ray
line emission\footnote{$J^{\pi}=2^+ \to 0^+$ transition  
in $^{44}Ca$ reached from $^{44}Sc$ following electron
capture.} from Cas A with the Compton Gamma Ray Observatory
was the first direct proof of synthesis of this short lived,  freshly made radio-isotope
in SNe. Renaud et al \cite{ren06} using the INTEGRAL spacecraft recently reported  the detection of
both 67.9 and 78.4 keV gamma-ray lines of $^{44}Sc$ in Cas A. 
There was a clear separation of the two lines due to an improved detection of the hard X-ray 
continuum up to 100 keV. The line flux of $(2.5 \pm 0.3) \times 10^{-5} \; \rm cm^{-2} \; s^{-1}$
leads to a tightly constrained $^{44}Ti$ mass of 
$(1.6^{+0.6}_{-0.3}) \times 10^{-4} M_{\odot}$). 
This is actually high compared to the predictions of the standard models \cite{Woo95}, 
\cite{thi96}, or their improved versions \cite{rau02}, \cite{lim03}. Since the production
of $^{44}Ti$ is sensitive to the explosion energy and asymmetries and Cas A is known to be
asymmetric and energetic ($2\times 10^{51}$ erg instead of the standard $1\times 10^{51}$ erg),
this could be a factor in its apparent overproduction compared to models.
At the same time, the recent revision of the  $^{40}Ca(\alpha,\gamma)^{44}Ti$ reaction
rate \cite{nas06} has led to an increase of $^{44}Ti$ production by a factor of $\sim 2$.

Live radioactive isotopes freshly synthesized in the explosion were detected from SN 1987A
by directly detecting 1238 keV and 847 keV gamma-ray lines \cite{mat88}, which provided detailed 
diagnostics of the nuclear burning conditions and the explosion dynamics. Another radio-nuclide
$^{57}Co$, which is a decay product of $^{57}Ni$ made in the supernova explosion, was also
detected directly by measuring gamma-ray lines \cite{kur92}.
The calculated gamma-ray light curves for 847 keV and 1238 keV and other lines \cite{nom91}
could be made consistent with the Solar Maximum Mission (SMM) measurement only if the
$^{56}Co$ was mixed up to a mass coordinate of $M_r \sim 13.5 M_{\odot}$. The calculated
fluxes were still slightly smaller than the observed fluxes at $t \sim 200 d$, but this
discrepancy could be removed by taking account of the effect of clumpiness in the ejecta
which is more significant in the earlier phases. Also, the observed flux ratio between the two 
gamma-ray lines at 847 keV and 1238 keV is close to unity at early stages because of the
smaller cross section for the 1238 keV lines than for 847 keV. As the column depth of the
overlying matter decreased due to dilution with time in an expanding envelope, the observed flux
ratio approached the expected value of 0.68 based on branching ratios \cite{nom91}.

X-ray fluorescence spectrometry technique has been used by archaeologists to determine the major 
and trace elements in ceramics, glass etc. Here,
a sample is irradiated with X-rays and the wavelengths of the released (fluorescent) 
X-rays are measured. Since different elements have characteristic
wavelengths, they can be identified and their concentrations estimated from the intensity of the 
released X-rays. Trace elements analysis may, for example, help in identifying the (geographical)
source of a material. A similar method has been proposed \cite{lei02} and attempted
for SN 1987A \cite{lei06}. It utilizes electron capture decays of freshly synthesized
radio-nuclei. Following the electron capture, the K-shell vacancies are in most cases
filled in by downward transitions from other bound shells. As the fluorescence X-ray yields
are well known, measuring the X-ray line fluxes can then estimate the number of freshly
synthesized nuclei. For example, attempts were made to detect the 5.9 keV K$\alpha$ line
from the stable nucleus $^{55}Mn$
in the Chandra X-ray spectrum of SN 1987A, which is due to the decay of 
radioactive  $^{55}Fe$. Both $^{55}Fe (\tau_{1/2} = 2.7 \; \rm yr)$ and $^{55}Co 
(\tau_{1/2} = 18 \; \rm hr)$ are produced in explosive, incomplete
Si burning as well as in normal freezeout of nuclear statistical equilibrium, in the inner ejecta
of core collapse supernovae. However, no evidence of the 5.9 keV line emission from Mn could be
found in 400 ks of Chandra ACIS data and the upper limit to the mean flux was 
$< 3 \times 10^{-7} \; \rm cm^{-2} s^{-1}$.
Rauscher et al \cite{rau02} calculated the ejected mass of $A=55$ radioactive nuclei
to be $7.7 \times 10^{-4} \; M_{\odot}$ for $20 M_{\odot}$ models of which most was $^{55}Co$.
If only about half this mass of $^{55}Fe$ were ejected, the reduced flux would be consistent
with the observed upper limit. On the other hand the, even if the total mass inside were
as much as $1 \times 10^{-3} \; M_{\odot}$, but the $^{55}Fe$ abundance was zero outside
the radial velocity shells at $1500 \; \rm km s^{-1}$, the line flux would be still consistent
with data, as at late times the emerging flux depends sensitively on the presence of $^{55}Fe$
in the outer zones.

\subsection{Other X-ray supernovae}

\begin{figure}[htb]
\centering
\vspace*{0.5cm}
\includegraphics[height=6cm]{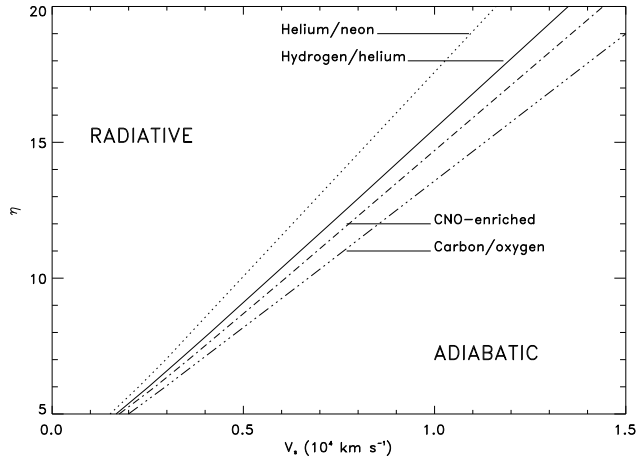}
\caption[]{
The border between radiative and adiabatic regions of the shocks in supernovae as a function of
shock velocity and ejecta density profile index $\eta$ ($\rho \propto v^{-\eta}$). The boundaries
depend upon the composition of the ejecta through which the shock may be propagating as also the
mass loss parameter of the progenitor stellar wind and the ejecta velocity scale
(Fig. courtesy 	T. Nymark \cite{nym09}).
}
\label{fig: shock-regimes}
\end{figure}

Apart from SN 1987A, which occurred close by and was therefore easy to detect,
there are supernovae with relatively high intrinsic X-ray luminosity. SN 1993J
and SN 1995N are two of them and they are
at the high end of the X-ray luminosity \cite{zim03}, \cite{fox00}, which makes it 
suitable for study in the X-ray wave bands even at late stages \cite{cha09}.
Nymark et al \cite{nym06} have shown that SNe with strong X-ray 
emission are likely to have radiative shocks and that in these shocks a large
range of temperatures contribute to the spectrum. The cooling of the shock
also affects the hydrodynamics of the flow structure and the volume of the
emitting gas and thus the total luminosity from the interaction region.
A shock is radiative depending on whether the cooling time scale $t_{cool}$
of the shock is short compared to its expansion time $t$. Since the cooling function 
(from electron bremsstrahlung or free-free scattering and X-ray
line emission from bound-bound transitions of ionized atomic species) is
a function of the shock temperature this condition translates
to a condition on the shock temperature for radiative vs. 
adiabatic shock. The boundary of the two regimes
depends upon the shock velocity $V_s$ and the ejecta density profile
(given by the index $\eta$ in $\rho \propto v^{-\eta}$) and the separation
in turn depends upon the chemical composition of the ejecta (see Fig.
\ref{fig: shock-regimes} reproduced from \cite{nym09}).
For a given shock velocity (and a composition and mass loss parameter of the progenitor 
stellar wind and the ejecta velocity scale), the shock is likely to be radiative
for steeper density gradients.

\begin{figure}[htb]
\centering
\includegraphics[height=7.5cm,width=11cm]{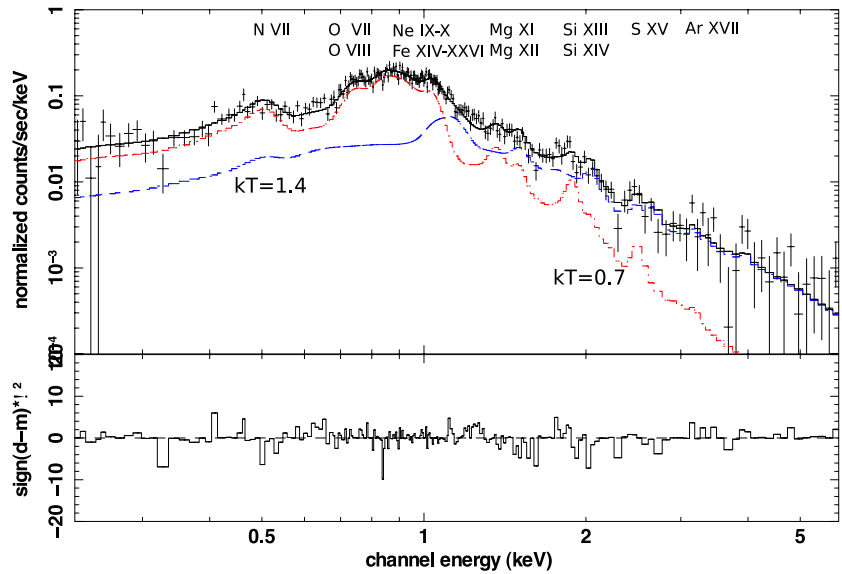}
\caption[]{
XMM spectrum of SN 1993J (shown by data-points with crosses from April 2001). The data
is fitted with an adiabatic shock at $kT_{rev} = 2.1 \; \rm keV$ (long dashed in blue)
and a radiative shock at $kT_{rev} = 1.0 \; \rm keV$ (short dashed in red) for a
He/N dominated composition from model s15s7b \cite{woo94}. 
(Fig. courtesy  T. Nymark \cite{nym09}).
}
\label{fig: xmm-sn1993j-spec}
\end{figure}

As already mentioned, the spectroscopic type of SN 1993J from a type II-like SN at
early times due to detected Balmer lines soon weakened and it came to resemble a type Ib SN,
leading to a reclassification as a type IIb SN \cite{fil93}. This  transition is best explained
if the progenitor star prior to its explosion, had lost its hydrogen envelope due to 
interaction with a binary companion \cite{ray93}, \cite{nom93}, \cite{pod93}, \cite{woo94},
\cite{mau04}.  If the H-envelope
was indeed already thin, then the reverse shock would transit the H-rich region quickly, and soon
the emission may be dominated by material dredged up from the nuclear processed parts of the star
in the interior. In the 4H47 model of Shigeyama et al \cite{shig94}, by 2600
days after explosion, the hydrogen-rich
envelope had $\sim 0.47 M_{\odot}$. Inside this envelope is the outer part of
the He-zone which is enriched in N (about $\sim 0.3 M_{\odot}$), while its inner part is C-rich
(has $\sim 1.5 M_{\odot}$). If the circumstellar medium or the ejecta is clumpy, they may
be hit by the reverse shock obliquely and the shock in the clump will be slower and the
shock temperature lower. The presence of both hard X-rays and optical emission  from the shocked
CSM shows the importance of a clumpy CSM, for example in SN 1987A where a range of shock
speeds are necessary to explain the observed spectrum \cite{zhe06}. SN 1993J is not as bright 
as SN 1987A or Cas A due to its larger distance, and both Chandra and XMM-Newton data is
comparatively of a moderate signal to noise, at the low (CCD) resolution. 
The XMM spectrum of SN 1993J is dominated by
the Fe L emission at 0.7 - 1.0 keV and is blended with strong Ne IX-X emission. Above 1.0 keV, 
emission lines which may be present are from Mg XI-XII, Si XIII-XIV, as well as S XVI.
Nymark et al find that the spectrum of SN 1993J is fit best by a combination of an adiabatic
shock ($kT_{rev} = 2.1 \; \rm keV$) propagating through a zone with CNO burning products, with
He dominant and N and Ne being the most abundant metals, and a second radiative shock
($kT_{rev} = 1.0 \; \rm keV$). The latter could have been caused by instabilities at the
reverse shock front or by a clumpy CSM. The XMM spectrum of SN 1993J and the model fits used
\cite{nym09} are shown in Fig \ref{fig: xmm-sn1993j-spec}.  

SN 1995N is a type IIn supernova, a type 
that shows unusual optical characteristics and
spans a very broad range of photometric
properties such as decline rates at late times \cite{fil97}. 
It is likely that these differences are related to
their progenitor's structure, mass, composition as well as 
the composition and the density profile of the CSM \cite{li02}.
These supernovae show the presence of strong, narrow Balmer line
emission on top of the broader emission lines in their early spectra.
The narrow emission lines may originate
in the dense and ionized circumstellar (CS) gas \cite{hen87,fil91b}.
The presence of strong H$\alpha$ emission line, the high bolometric
luminosity and the broad H$\alpha$ emission base 
powered by the interaction of the supernova shock with the CSM, all
point towards a very dense circumstellar environment \cite{chu03}.
When the stellar core collapses and explodes, the supernova lights up 
the slow-moving gas into narrow emission lines leading to the type IIn
supernova classification (n for narrow emission line).
This interaction of the supernova shock with
the dense CS gas is indicated by strong radio and X-ray emission
detected from  several type IIn supernovae and in particular SN 1995N.

Chandra ACIS observations of SN 1995N show a  best-fit line energy at 1.02 keV
which is ascribed to Ne~X \cite{Cha05}.
There is also the possibility of another line with best-fit line energy at 0.85 keV with  
identification as Ne~IX. 
The ionization potentials of all ionized Neon 
species upto Ne~VIII are less than  240 eV, whereas those of
Ne~IX and Ne~X are above 1195 eV. Hence at  
temperatures found by the Chandra observations, the predominant 
species of Neon are expected to be Ne~X and Ne~XI.
The mass of Neon 
was estimated to be
about $5\times 10^{-3}M_{\odot}-
1\times10^{-2}M_{\odot}$. This is consistent with Neon being 
in the Helium layer for a $15 M_{\odot}$ star
where it was co-synthesized with C, N, and He. 
For other stellar masses and other composition zones, the
required Neon mass is much larger than
observed. Therefore, Neon in the Helium core
of the $15 M_{\odot}$ star is the probable site where it was co-synthesized
with C, N.

\section*{Conclusions}
Much insight about how stars burn nuclear fuel, evolve and ultimately explode
has been obtained by calculations and 
computer simulations. Theoretical developments have been augmented 
by crucial astronomical observations over decades in 
all bands of electromagnetic radiation and through other signal channels like
neutrinos. Most important have been the wider availability of powerful telescopes
both from the ground and space platforms and state of the art
computers. 
Crucial inputs to the field of nuclear astrophysics are 
also coming from laboratory experiments involving radioactive ion beams (RIB) 
and intense beams of energetic nucleons and nuclei. Short lived nuclei 
can only be studied close to their sites of formation in the laboratory before
they decay. Such facilities  will further advance the future of nuclear astrophysics.

\section*{Acknowledgments}
I thank Aruna Goswami and the organizers of the School
for the invitation to Kodaikanal Observatory.
Discussions with Katherina Lodders and Bruce Fegley at the School and
Claus Rolfs and Richard McCray in other meetings are thankfully
acknowledged. I thank Poonam Chandra and Firoza Sutaria for comments on this 
manuscript and for long term collaboration.  
Nuclear astrophysics research at Tata Institute is a part of the 
Plan Project: 11P-409.

\printindex
\end{document}